\documentclass[12pt,citesort]{article}

\newcommand\Black{}

\usepackage{axodraw}
\usepackage{color}
\usepackage[dvips]{graphicx}
\usepackage{graphicx}
\usepackage{amsmath,amssymb}

\newcommand{\half}{ {\scriptstyle \frac{1}{2} } }
\newcommand{\Half}{ {\frac{1}{2} } }

\newcommand\be{\begin{equation}}
\newcommand\ee{\end{equation}}
\newcommand{\bel}[1]{\begin{equation}\label{#1}}
\newcommand\bea{\begin{eqnarray}}
\newcommand\eea{\end{eqnarray}}
\newcommand{\bdm}{\begin{displaymath}}
\newcommand{\edm}{\end{displaymath}}
\newcommand\nn{ \nonumber\\}

\newcommand{\tbox}[1]{\qquad \mbox{#1} \qquad}

\newcommand{\dd}[1]{\partial_{#1}}
\def\dd{\partial}
\newcommand{\<}{\langle}
\renewcommand{\>}{\rangle}

\makeatletter
\@addtoreset{equation}{section}
\makeatother

\title{On The Pomeron at Large 't Hooft Coupling}

\author{ Richard  C. Brower\footnote{Physics Department,
Boston University, Boston, MA 02215},\
Matthew J. Strassler\footnote{Department of Physics and Astronomy, Rutgers
University, Piscataway,  NJ 08854}
\ and  Chung-I Tan\footnote{Physics Department, Brown University,
Providence, RI 02912}\\ \\
E-mail: brower@bu.edu,strassler@physics.rutgers.edu,tan@het.brown.edu
}

\begin{document}
\maketitle

\begin{abstract}
We begin the process of unitarizing the Pomeron at large 't Hooft
coupling.  We do so first in the conformal regime, which applies to
good accuracy to a number of real and toy problems in QCD.  We rewrite
the conformal Pomeron in the $J$-plane and transverse position space,
and then work out the eikonal approximation to multiple Pomeron
exchange.  This is done in the context of a more general treatment of
the complex $J$-plane and the geometric consequences of conformal
invariance.  The methods required are direct generalizations of our
previous work on single Pomeron exchange and on multiple graviton
exchange in AdS space, and should form a starting point for other
investigations.  We consider unitarity and saturation in the conformal
regime, noting elastic and absorptive effects, and exploring where
different processes dominate.  Our methods extend to confining
theories and we briefly consider the Pomeron kernel in this context.
Though there is important model dependence that requires detailed
consideration, the eikonal approximation indicates that the
Froissart bound is generically both satisfied and saturated.
\end{abstract}

\newpage
\setcounter{tocdepth}{3}
\tableofcontents

\newpage

\section{Introduction}
\label{sec:intro}

Recently there has been considerable effort --- much of it using
gauge/string duality --- toward understanding high energy hadronic
scattering in gauge theories.  The conceptual and experimental
importance of this problem is widely known, as is its difficulty.
Here we take a step forward that will allow a number of interesting
problems to be addressed.

The goal of this paper is to continue developing methods for
implementing unitarity at finite 't Hooft coupling $\lambda$.  In a
wide regime of high energy scattering, Pomeron exchange is believed to
be the dominant, universal process
~\cite{Chew:1961ev,Gribov:1961fr,Lipatov:1976zz,Kuraev:1977fs,BL,Capella:1984tj,DL}.
 The Pomeron kernel with a leading
$J$-plane singularity at $j_0>1$ contributes to cross sections a growing
power $s^{j_0-1}$ (up to logarithms) in violation of unitarity.  The
eikonal approximation is an order-by-order summation of the leading
$(s^{j_0-1}/N^2)^n$ contribution to the $n$-Pomeron exchange
process; here $N$ is the number of colors.  
This approximation can
only be valid in limited regimes, where the scattering angle is small.
But understanding it represents a first step in satisfying the
non-perturbative unitarization of high energy scattering
~\cite{Giudice:2001ce,Giddings:2002cd,Giddings:2007bw}.

A smaller step toward introducing unitarity corrections was made
recently in the eikonal approximation for pure $AdS_5$
gravity,~\cite{Cor,Brower:2007qh}; earlier related work includes
~\cite{Cornalba:2006xk,Cornalba:2006xm}.
This limit represents taking $\lambda \rightarrow
\infty$ prior to considering large $s$.
The power $j_0$ is exactly 2 in this
case.

Here we keep $\lambda$ large but finite, and consider the richer
question of Pomeron exchange with $j_0<2$, which is more relevant for
QCD-like theories which have $j_0$ considerably less than $2$.  We
mainly present results for conformal field theories, but our results
can easily be generalized to nonconformal theories, which we consider
here only briefly.  More difficult generalizations will involve including
nonlinear effects and/or moving beyond the eikonal approximation.

The basic concepts involving high-energy hadron scattering in
gauge/string duality have emerged in stages.  It was shown in
\cite{Polchinski:2001tt} that in exclusive hadron scattering, the dual
string theory amplitudes, which in flat space are exponentially
suppressed at wide angle, instead give the power laws that are
expected in a gauge theory.  Other related work includes
~\cite{Brower:2002er,brodsky}.  
It was also argued that at large $s$
and small $t$ that the classic Regge form of the scattering amplitude,
varying as $s^{t/t_0}$, is found in certain kinematic regimes.
Next, deep inelastic scattering was studied \cite{DIS}.  The moderate
$x$ regime was shown to be quite different from that of weak-coupling
gauge theories due to much more rapid evolution of structure functions
with $q^2$.  The classical-twist-two operators develop large anomalous
dimensions and (at finite but large $N$) become subleading compared to
double-trace, higher twist operators.  At small $x$, by contrast, the
physics is more similar to that of weak coupling, with a large growth
in the structure functions controlled by Regge physics.  The correct
equations for the string-theory realization of the gauge-theory
Pomeron were identified, along with the fact that the growth of the
structure functions is controlled by a power $j_0-1$ with $j_0$ very
close to, but less than, 2.  After several interesting attempts using
other
methods~\cite{Janik:1999zk,Janik:2000pp,Andreev:2004sy,Nastase:2005bk},
it was demonstrated in~\cite{Brower:2006ea} that the Pomeron equations
of~\cite{DIS} could be easily solved and interpreted.  The Pomeron was
identified as a well-defined feature of the curved-space string
theory.  Its mathematical form in the conformal region of a gauge
theory --- the ``hard'' Pomeron --- was shown to share many of the
feature of the BFKL Pomeron for weak-coupling gauge theories.  The
Pomeron in the confining region was shown to have the features one
would expect from QCD; running trajectories with bound states at
integer $j$.  Moreover, the ``hard'' and ``soft'' Pomerons were shown
to be a single, unified object, as conjectured in
\cite{levintan,Bondarenko:2003xb}.  With this understanding of the
Pomeron, any computation involving single-Pomeron exchange in a
confining gauge theory can now, at least in principle, be carried out.

However, as we mentioned above, single-Pomeron exchange is only
appropriate in limited regimes, since it violates unitarity at
high energy.  We now turn to the question of summing multiple Pomeron
exchange where the scattering is sufficiently weak.  As a preliminary,
let us recall a result of \cite{Brower:2006ea}.  In a conformal field
theory, the Pomeron exchange kernel in the gauge theory can be
represented in the dual string theory through a kernel ${\cal K}$, a
function of $s,t$ and two bulk coordinates $z, z'$.\footnote{Here the
metric on the Poincare' patch of
$AdS_5$ is
\be ds^2= \frac{R^2}{z^2}\left[\eta_{\mu\nu} dx^\mu dx^\nu
+ dz^2\right]
\ee
}
This kernel is akin to a propagator for the
Pomeron.  At $t = 0$ it has the very simple form,
\be
{\rm Im}\ {\cal K}(s,0,z,z') \simeq
\frac{s^{j_0}}{\sqrt{ \pi  {\cal D}\ln s}}
e^{-({\ln z - \ln z'})^2/{\cal D}\ln s}
\label{eq:ImKzero}
\ee
where $j_0=2-2/\sqrt{\lambda }$ and ${\cal D}= 2/\sqrt{\lambda}$.  This is
strikingly similar to the weak BFKL
kernel~\cite{Lipatov:1976zz,Kuraev:1977fs,BL},
\be
{\rm  Im}\ {\cal K} (p_\perp,p'_\perp,s)\approx
{s^{{j_0} } \over \sqrt{{
\pi{\cal D} }\ln s}}
e^{-(\ln p'_\perp-\ln p_\perp)^2/
{\cal D}\ln s}
\ee
with $j_0 = 1 +(4\ln 2 / \pi) \alpha N$, ${\cal D} = (14\zeta(3) /\pi
) \alpha N$, $\alpha= g^2_{YM}/4\pi$.  This correspondence identifies
diffusion in virtuality (or $\log p_\perp^2$ for the off-shell gluons)
with diffusion in the radial co-ordinate $\log z^2$ in the dual
$AdS_5$ space.\footnote{In \cite{Brower:2006ea} the diffusion was
taken with respect to $\log z$, or $\log p_\perp$, while standard
conventions, to which we adhere in this paper, take diffusion in the
variable $\log z^2$ or $\log p_\perp^2$.  Consequently the diffusion
constant used here is normalized differently, compared to
\cite{Brower:2006ea}, by a factor of 4.}  For $t\neq 0$ the form is
somewhat more complicated, and is given below.

Our task will be to first
reconsider the Pomeron kernel of \cite{Brower:2006ea} in the conformal
regime.  We will rewrite it in impact parameter space and in the $J$-plane,
which greatly simplifies its form.  In particular, as is the case for
graviton exchange \cite{Cor,Brower:2007qh}, it involves an
$AdS_3$ scalar Green's function.  We then use this answer to construct
the eikonal approximation to the full amplitude, as we did in
\cite{Brower:2007qh}.  In doing so we include an infinite ladder of
Pomeron exchanges, but neglect all non-linear Pomeron interactions.
Of course this approximation is only valid in limited regimes, but we
will not address the region of validity here.  Within the regime in
which it applies, the eikonal amplitude satisfies a form of bulk
unitarity and exhibits both elastic and absorptive parts.

Our main results are the following.  We begin with a conformal large-$N$
theory.  For definiteness, consider adding a massive probe to such a
theory, whose effects on the dynamics are subleading in $1/N$, such as a
massive quark in the fundamental representation.  Then
consider $2\to 2$ high-energy scattering
of gauge-neutral states associated to the probe
(such as massive quarkonium states) at fixed $s,t$.  The string
description of this scattering
depends not only on $s,t$ but also on the bulk location $z$ where the
scattering occurs.  In fact this is not enough; in general the
scattering is not local, and could depend on the $z$ locations of all
four strings.  However, there may be some regions of $z, z'$, and
impact parameter $b$ (where $b\equiv x_\perp-x'_\perp$ is the distance
in the two Minkowski space coordinates transverse to the motion),
where the eikonal regime is valid.  Let us separate the amplitude
$A_{2\to2}(s,t)$ into the region in position space variables where the
eikonal approximation is valid --- the ``eikonal region'' ${\cal E}$
--- and the regime where it is not.  The contribution of the eikonal
region to the amplitude --- to which the amplitude in the non-eikonal
region, if any, must be added --- is
\be
- 2i s \int_{{\cal E}}\  d^2b \
 dz \ dz'\ P_{13}(z) P_{24}(z') e^{- ib^\perp q_\perp} \left [
e^{i\chi(s,x^\perp - x'^\perp, z,z')} - 1\right]
\label{eq:adsiek1}
\ee
(Here the scattering is of initial states $1,2$ to final states $3,4$.)
The wave functions $\Phi_i$ for the
scattered states appear in pairs, evaluated at $z$ for the {right}-moving
states and at $z'$ for the {left}-moving states:
\be \label{eq:overlapfunction1}
P_{13}(z) = (z/R)^2\sqrt{g(z)}  \Phi_1(z) \Phi_3(z) \tbox{and} P_{24}(z')
= (z'/R)^2\sqrt{g(z')}  \Phi_2(z') \Phi_4(z')
\ee
The eikonal kernel $\chi$ is then a function of $b, z, z'$ as well as $s$.
This form of the amplitude is identical to that found for the eikonal
approximation for graviton scattering
in $AdS_5$ space, except that now the function $\chi$ is proportional
not to the graviton propagator (projected onto $AdS_3$) but to the Pomeron
exchange kernel:
\bel{AdSpomeronchi1}
\chi(s,x^\perp- x'^\perp,z,z')=
  \frac{ \kappa_5^2 R }{ 2(zz')^2 s}  {\cal K}(s,x^\perp - x'^\perp,z,z')
\ee
where $\kappa_5$ is the gravitational coupling constant in $AdS_5$. In
the following, instead of $\kappa_5^2$, we will often use instead a
dimensionless coupling $g_0^2= \kappa_5^2/R^3\sim 1/N^2$.

\begin{figure}[h]
\begin{center}
\includegraphics[width = 3.7in]{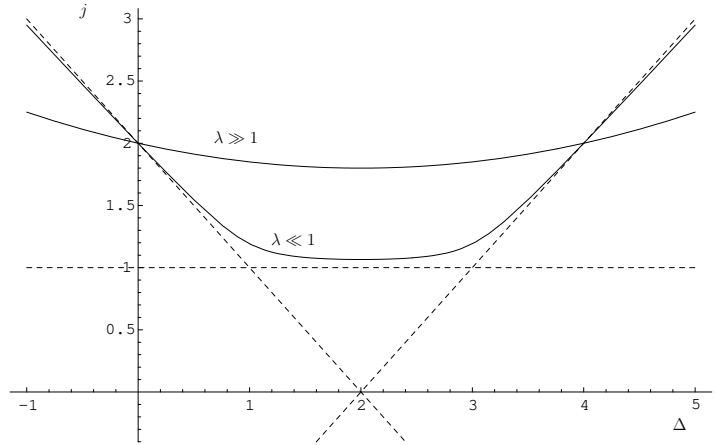}
\caption{Schematic form of the $\Delta-j$ relation for $\lambda\ll 1$
and $\lambda\gg 1$.  The dashed lines show the $\lambda =0$ DGLAP
branch (slope 1), BFKL branch (slope 0), and inverted DGLAP branch
(slope $-1$).  Note that the curves pass through the points (4,2) and
(0,2) where the anomalous dimension must vanish.  This curve is often
plotted in terms of $\Delta - j$ instead of $\Delta$, but this obscures
the inversion symmetry $\Delta \to 4-\Delta$.}
\label{fig:BFKLDGLAP}
\end{center}
\end{figure}

The kernel, as a
function of $s$ and transverse positions, is elegantly expressed
through an inverse Mellin transform
\be
{\cal K}(s,x^\perp-x'^\perp,z,z') = - \int \frac{dj}{2\pi i}
\left(\frac{{ \widehat s}^j+{(-\widehat s)}^j}{ \sin \pi j} \right) {\cal
K}(j,x^\perp-x'^\perp,z,z') \; .
\label{eq:pomeronkernel1}
\ee
where
\be
\widehat s\equiv z z' s
\ee
is dimensionless and
\be
{\cal K}(j,x^\perp - x'^\perp ,z,z')
= (z z' /R^4)G_3(j,v) \; ,
\label{eq:ads3kernel1}
\ee
Here $G_3(j,v)$ is the $AdS_3$ Green's function which has a simple closed
form,
 \be
G_3(j,v)= \frac{1}{4\pi } \frac{ \Big[1+v + \sqrt{v(2+v)}\Big]^{ (2 -
\Delta_+(j))}}{  \sqrt{v(2+v)}} \; .
\label{eq:adschordal}
\ee
It depends on the $AdS_3$ chordal distance,
 \be\label{eq:vdefn}
v= \frac{(x_\perp-x'_\perp)^2+(z-z')^2}{2 zz'}
\ee
and the $AdS_3$ conformal dimension, $\Delta_+(j)-1$, where
\be
\Delta_+(j) =2+\sqrt{4 + 2\sqrt \lambda(j-2)}= 2 + \sqrt {2\sqrt
\lambda(j-j_0) }
\label{eq:Delta4}
\ee
sets the dimension $\Delta$ as a function of spin $j$ for the
BFKL/DGLAP operators.
The analytic continuation from DGLAP to
BFKL operators has been discussed at weak coupling for some time
\cite{Jaroszewicz:1982gr,Lipatov:1996ts,Kotikov:2000pm,Kotikov:2002ab}.  Recently, it was conjectured to be exact at weak coupling
in ${\cal N}=4$ Yang-Mills theory \cite{klv5}.  The
demonstration of
this relationship in all large-$\lambda$ conformal theories, and the
derivation of the formula (\ref{eq:Delta4}), is given in section 3 of
\cite{Brower:2006ea}, where
the existence of the single function
$\Delta_+(j)$ with $j=j_0$ at $\Delta=2$ (the BFKL exponent) and $j=2$ at
$\Delta = 4$ (for the energy-momentum tensor, the first DGLAP
operator) was demonstrated.
For clarity, we reproduce
Fig.~\ref{fig:BFKLDGLAP} from \cite{Brower:2006ea} showing the
essential form of this function for large and small
$\lambda$.

\begin{figure}[h]
\begin{center}
\includegraphics[width = 3.7in]{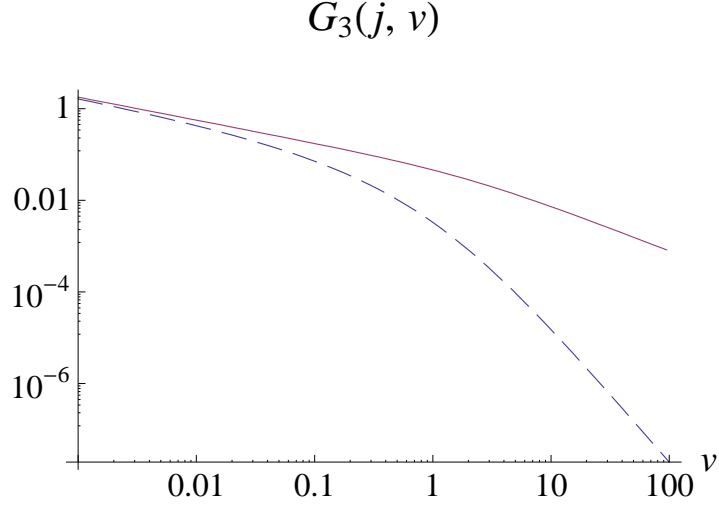}
\caption{The function $G_3(j,v)$ for $j=j_0$ (the solid line for the Pomeron, with
$\Delta=2$) and for $j=2$ (the dashed line for the graviton, with $\Delta=4$).  Note
$G_3(j,v)\sim 1/\sqrt v$ for $v\ll 1$ and $\sim v^{1-\Delta_+(j)}$ for
$v\gg 1$. Thus the two functions have the same small-$v$ behavior, but
the graviton falls off faster than the Pomeron.}\label{fig:G3}
\end{center}
\end{figure}

The function $G_3$ is shown in Fig.~\ref{fig:G3} for
the Pomeron and graviton.
Naturally it becomes large
as $v\to 0$; there the scattering is head-on in the bulk, $\chi$ becomes
large and strongly varying, and the eikonal approximation will break
down.  Conversely at large $v$ the eikonal approximation will be good;
note this includes both large $b$ compared to $z$ and $z'$ (where the
scattered objects are at large impact parameter compared to their
size) and at large $z-z'$ for fixed $z$ and $z'$ (where their sizes are
mismatched.)

We obtain the above form of the kernel as a Fourier transform
of the conformal Pomeron
\be
{\cal K}(j,x^\perp - x'^\perp ,z,z')
= \int \frac{d^2q_\perp}{(2\pi)^2}  e^{i q_\perp( x^\perp - x'^\perp)}
{\cal K}(j,-q_\perp^2 ,z,z')  \label{eq:2dFT1}
\ee
which was found in \cite{Brower:2006ea} to be
\be
{\cal K}(j, t,z,z')=  \frac{(zz')^2}{\pi^2R^4}\int_{-\infty}^\infty d\nu
(\nu\sinh \pi \nu) \frac{ K_{i\nu}(qz)K_{-i\nu}(qz')}{ \nu^2  + (2\sqrt
\lambda) (j- j_0)}\; .
 \label{eq:jPomeron1}
\ee
This in turn is a Mellin transform of the imaginary part of the kernel
as a function of $s,t,z, z'$
\be
{\cal K}(j,t,z,z')= \int^\infty_{0} d\hat s \; ( \widehat s)^{-j-1} \;\;{\rm
Im} \;\;{\cal K}( s, t, z,z')
\label{eq:MellinofImK1}
\ee
where again $\hat s\equiv z z' s$.
This strong coupling kernel~\cite{Brower:2006ea} is a momentum
space Green's function propagating in $AdS_5$, satisfying
\be
[  -    z^5 \dd_z z^{-3} \dd_z  -  z^2 t +2\sqrt{\lambda}(j-2) ]{\cal
K}(j,t,z,z')= R^{-4} z^5\delta(z-z') \ .
\label{eq:ads5qDE1}
\ee
It is also convenient to consider the spectral decomposition
of the kernel with respect to $t$,
\be
{\cal K}(j,t,z,z') = \frac{(zz')^2}{2R^4}\int_0^\infty  dk^2 \frac
{J_{\widetilde \Delta(j)} (zk) J_{\widetilde \Delta(j)}(z'k)}{ k^2 -t 
-i\epsilon }
\ .
\label{eq:tspectral1}
\ee
Here $\widetilde \Delta (j) \equiv \Delta_+(j)-2$.

Note finally that the full kernel, rather than just its imaginary part,
can be reconstructed from the above expressions through
\be
{\cal K}(s,t,z.z') = - \int \frac{dj}{2\pi i}  \left(\frac{{\widehat
s}^j+{(-\widehat s)}^j}{ \sin \pi j} \right)  {\cal K}(j,t,z,z') \ .
\label{eq:AdSinvMellin1}
\ee
Here, as in the flat-space string theory, the contour of integration
is to the left of the poles at non-negative integers from the $1/\sin \pi
j$ factor and to the right of singularities of ${\cal K}(j,t,z,z')$.

After deriving these results in Sections 2 and 3, we discuss some
interesting features of the single Pomeron exchange kernel in Sec.~4.
We examine why the $AdS_3$ Green's
functions emerge in the form of the kernel, and the algebraic
structure which underlies our formula for $\Delta_+(j)$. Next, we
examine the high-energy behavior of the kernel as a function of $s$
and $b$. As $s\to
\infty$, with $\lambda$ fixed, the Pomeron exchange is dominant, with
\be
\chi\sim e^{i(1-j_0/2)\pi} \widehat s^{j_0-1}G_3(j=j_0,v) \sim \frac{
e^{i(1-j_0/2)\pi} \hat s^{j_0-1}}{\sqrt{v(2+v)}} \ \ (s\to\infty) \ .
\ee
Note that the overall phase of the
kernel is $\exp[i(1-j_0/2) \pi]$, independent of $b,z,z'$.
However, at large fixed $s$ and
$\lambda\to \infty$, we recover the graviton exchange kernel of
\cite{Cor,Brower:2007qh},
\be
\chi\sim \widehat s G_3(j=2,v) \sim \frac{\widehat
s}{[1+v+\sqrt{v(2+v)}]^2\sqrt{v(2+v)}}
\ \ \ \ (\lambda\to
\infty) \ .
\ee
where $G_3(2,v)$, called $G_3(v)$ in \cite{Brower:2007qh},
is the dimensionless scalar
propagator for a particle of mass $\sqrt{3}/R$ in an $AdS_3$ space of
curvature radius $R$.

In Sec.~5 we turn to the features of the eikonal sum of multiple
Pomeron exchanges.  We discuss the physics of the non-trivial phase of
the Pomeron and how the effects of absorption are distributed over the
bulk.  Finally we reinterpret our results as those of a multi-channel
eikonalization in the gauge-theory, consider how amplitudes for
scattering a particle off a Pomeron are embedded in our results, and
note some relations of our results with the string theory eikonal
approximation \cite{Amati:1987wq, Amati:1987uf}.  In particular we
note that within the eikonal approximation, well known to be
instantaneous in light-cone time, the scattering process acts
on each bit of string independently, giving it its own eikonal phase.

Finally, in Sec.~6 we turn to the question of the saturation of the
unitarity bound in various contexts.\footnote{After the initial
version of this paper was released, we noticed an error which
invalidated some of the results of Sec.~6.  Our corrected results in
the conformal limit now turn out to be reasonably consistent with
those of \cite{HIM}, which studied a different but related problem.}
We do this first for the bulk amplitude in a conformal theory.  Next
we consider briefly confining backgrounds, where the continuum
spectrum in $t$ at fixed spin-$j$ in Eq.~(\ref{eq:tspectral1}) becomes
a discrete sum over Regge trajectories.  Our general methods are still
applicable, though the technical difficulties and model-dependence are
much greater, and our results are very limited.  But we will argue
from the eikonal approximation that the cross section appears
generically to be proportional to $(\log s)^2$, both satisfying and
saturating the Froissart bound.

\newpage

\section{Overview of Regge behavior in string theory}
\label{sec:regge}

String theory was invented largely to accommodate
two phenomenological features of hadronic scattering
related to Regge theory. The first is the existence of
narrow resonances of higher and higher spin apparently lying 
on  almost linear ``trajectories'' ($j \sim m^2_j + \mbox{const}$).
The second is the Regge asymptotic limit for high energy scattering at
fixed momentum transfer,
\be
A(s,t) \sim s^{j_0 + \alpha' t}
\ee
where the trajectory function $\alpha(t) = j_0 + \alpha't$ is an
extrapolation to $t<0$ of the linear relation $\alpha(m^2_j) \simeq
j$, for $j>0$.  While this has proven to be an oversimplification, the
proper relationship between singularities in the complex $J$-plane and
high energy scattering was thoroughly investigated in this
context. Indeed the multi-Regge behavior of the planar limit of flat
space string theory with exactly linear trajectories provides an
excellent pedagogical tool~\cite{Brower:1974yv}.  Consequently, before
we extend this analysis to the recently understood Regge limit for the
$AdS$ dual to gauge theories, it is useful to review the arguments
briefly in flat space.  Moreover, as we note below, the general Regge
framework introduced here is valid for any theory with sufficient
convergence at high energies.

\subsection{Role of  $J$-plane singularities in flat space string theory}
\label{sec:Jplane}

In flat space, the tree-level string scattering amplitude has a
meromorphic representation in the complex $J$-plane. The argument
proceeds as follows.  The high energy limit of tachyon scattering
amplitude in the closed string sector is
\be
A(s,t) = \int d^2w |w|^{-2 - \alpha(t)} |1 - w|^{-2 - \alpha(s)} \simeq 
 2 \pi \frac{\Gamma(- \alpha(t)/2)}{\Gamma(1+\alpha(t)/2)}  (e^{-i \pi/2} \alpha' s/4)^{\alpha(t)}  
\label{eq:ReggeTachyon}
\ee
where $\alpha(t) = 2 + \alpha't/2$. Since the original amplitude is
 crossing symmetric under the exchange $u \leftrightarrow s$, by  virtue of $s+t+u=4m^2$,   Eq.~(\ref{eq:ReggeTachyon}) may be rewritten as,
\be
A(s,t) \simeq - \beta(t) \left[\frac{(- \alpha' s)^{\alpha(t)} +  (- \alpha' u)^{\alpha(t)}}
{\sin \pi \alpha(t)} \right] 
\label{eq:leadingasymbeh}
\ee
to leading order at high energy, $u \simeq -s $, where the residue
function\footnote{For superstring graviton-graviton scattering, the
residue $\beta(t)$ has an extra factor of $(\alpha(t)/2)^2$ to remove
the tachyon~\cite{Amati:1987uf}.  In the open superstring sector the
residue function has the simpler form $\beta(t) =
\alpha(t)/\Gamma[-\alpha(t)]$.} is $\beta(t) = 2^{1- 2
\alpha(t)}\pi^2/ \Gamma^2(1 + \alpha/2)$. This form is preferable,
since the separation of the Regge contribution from the right-hand cut, $
(-\alpha' s)^{\alpha(t)}$, and the left-hand cut, $ (-\alpha' u)^{\alpha(t)}$,
obeys exact crossing symmetry.  
Moreover it can be shown that the full
string amplitude $A(s,t)$, for large $s$ away from the singularities on
the real axis, is given by a sum of powers in $s$, without $\log s$
corrections.  This implies there are poles but not cuts, and
hence meromorphy, in the complex $J$-plane.

Now let us return to the full amplitude to investigate the source of
Regge behavior in terms of a complex $J$-plane. To relate this to the
singularity structure of the complex $J$-plane requires two
steps. First, the amplitude $A(s,t)$ must be expressed as a dispersion
relations over the right-hand ($s>0$) and left-hand ($u>0$) cuts
(each of which is actually a series of delta functions in
tree-level string theory). Second, each contribution to the imaginary parts 
associated to the cuts, $A_s$ and $A_u$, must be separately transformed to the
$J$-plane by
\be
a_s(j,t) =\alpha'  \int^\infty_{0}  ds \; (\alpha' s)^{-j-1} A_s(s,t) \; ,
\ee
and similarly for $s \rightarrow u$ which for our crossing symmetric
amplitude (\ref{eq:ReggeTachyon}) implies $a_u(j,t) = a_s(j,t)$.
Assuming, for some fixed $t<0$, that $A_s(s,t)$ is zero for $s \in
[0,s_0]$, this is merely the Laplace transform in rapidity $y =
\ln(s/s_0)$, giving an analytic function in $j$, defined initially for
large enough ${\rm Re} \; j$.  The inverse Mellin transform is given
by the contour integral,
\be
A_s(s,t)  =  \int^{i\infty+J_0}_{- i \infty +J_0} \frac{dj}{2\pi i} \; (\alpha' s)^{j} 
\;  a_s(j,t) \; ,
\ee
choosing $J_0$ to the right of all singularities. This inversion
becomes clearer when viewed as a Fourier transform in ${\rm Im}\ j$.

For $t$ sufficiently negative such that $A(s,t) = 0(1/|s|)$, the
dispersion relation for $A(s,t)$, 
\be
A(s,t) = \int^\infty_{0} \frac{ds'}{\pi} \frac{A_s(s',t)}{s'-s- i\epsilon} + 
\int^\infty_{0} \frac{du'}{\pi} \frac{A_u(u',t)}{u'-u - i\epsilon}  
\label{eq:DR}
\ee
allows us to reconstruct the full amplitude~\footnote{There are a
  variety of closely related transforms that define the $J$-plane with
  identical leading singularities. For example from the $t$-channel
  partial wave expansion, one is lead to the Sommerfeld-Watson
  transform, $$A(s,t) = - \sum_{\eta = \pm1}\int\frac{dj}{2\pi i} \;
  (2 j+1) \; \frac{\eta + e^{-i\pi j}}{2 \sin \pi j } a^\eta(j,t) \;
  P(j,(s-u) /t) \; ,
$$ where  $\eta$ is referred to as the signature: $\eta = \pm 1$
for $C = \pm 1$ exchange respectively.}

\be
A(s,t) = - \int^{ i\infty+J_0}_{ -i\infty+J_0 } \frac{dj}{2\pi i} \; \frac{ (-  \alpha'  s)^{j}   a_s(j,t) + (-\alpha' u)^{j}  a_u(j,t) }{\sin \pi j} \;
\label{eq:Regge}
\ee
from the $J$-plane, where $J_0\simeq -1$.

As one increases $t$, poles in $j$ move to the right, and one must
distort the contour to stay to the right of the
singularity. Therefore, for general $t$, the contour should be to the
left of the pole in $1/\sin\pi j$ at $j=0$ and to the right of all
singularities of $a_s(j,t)$ and $a_u(j,t)$.  For example, 
in an amplitude with vacuum quantum numbers in the $t$-channel, the
leading pole is symmetric in $s\leftrightarrow u$ interchange,
\bel{flatpartialwave}
a_s(j,t) \simeq \frac{\beta(t) }{j - \alpha(t)} \quad , \quad  a_u(j,t) \simeq \frac{\beta(t) }{j - \alpha(t)}
\ee 
with positive charge conjugation $C=+1$. This leads to an 
amplitude,
\be
 A(s,t) \simeq -\frac{(1 + e^{-i \pi\alpha(t)}) \beta(t) s^{\alpha(t)}}{\sin \pi \alpha(t)} \sim  \Gamma[-\alpha(t)/2] (e^{-i\pi/2} s)^{\alpha(t)}  
\ee
with ``positive signature'' in the language of Regge theory.  This
effect of the leading trajectory reproduces the leading Regge
approximation of our string amplitude, Eq.~(\ref{eq:leadingasymbeh}),
for all $t$.  Recall that in flat space closed string theory, the leading
trajectory, which contains the zero mass graviton at $j=2$, is the
analogue of the Pomeron in gauge theory.  

As an aside, we note that the leading negative charge conjugation
($C=-1$) contribution is odd under $s\leftrightarrow u$ interchange, giving a
negative signature factor $1 - e^{-i\pi\alpha(t)}$. This contribution,
in the context of weak coupling QCD, is the analogue of the BFKL
Pomeron referred to as the ``odderon'' with an odd number of gluons
exchanged in the $t$-channel. Both contributions are present in an
oriented close string exchange process.

The generality of the definition of an analytic $J$-plane should be
clear, in spite of our use of the planar closed string amplitude as a
convenient pedagogical example. In general, complete knowledge of the
$J$-plane singularity structure allows an {\it exact} representation
of the full amplitude, if the amplitude has the required convergence for
unsubtracted dispersion relations at sufficiently negative $t$.
Although Born terms in a perturbative field theory fail to satisfy
this dispersion relation constraint, it is generally believed that
full QCD does satisfy it, as well as a wide class of perturbative
string theories, order by order in $1/N$ or $g_s$.  The flat-space critical
superstring is the classic example, with an additional special feature
that the tree-level amplitude exhibits meromorphy in {\it both} energy
and the $J$-plane.~\footnote{The proof of meromorphy to the closed
string tachyon scattering amplitude (\ref{eq:ReggeTachyon}) is most
easily done by using a modified $J$-plane defined by the Beta
transform:
$$\widetilde a_s(j,t) = \int^{i\infty}_{-i \infty} ds A(s,t)
B(\alpha(s)/2 +1,j+1) = 2
\pi\frac{\Gamma[j-\alpha(t)]\Gamma[j+1]}{\Gamma[j+1-\alpha(t)/2]^2} \;
. $$ It can be shown that this implies that the Mellin transform is an
equivalent but less elegant meromorphic representation of the
$J$-plane.} However, one lesson from gauge/string duality is that the
gauge theory amplitudes dual to string theory in curved space need not
show meromorphy in the $J$-plane, even in the planar limit; for
instance this is illustrated by the BFKL singularity for the hard
Pomeron in large-$N$ conformal field theories, where conformal
invariance assures the presence of cuts in the $J$-plane. New branch
cuts in the $J$-plane also show up, for both flat space string theory
and gauge theory, at higher orders in $g_s$ or $1/N^2$, and thus 
in an
eikonal sum. Nevertheless, the knowledge of the $J$-plane
singularities can in principle allow a full reconstruction of the
full amplitudes.

\subsection{An Aside on Fixed Poles}
\label{sec:fixedpole}

Here we address a general issue which is useful later, but can be
omitted at a first reading.  In Secs.~\ref{sec:twopomcut} and \ref{sec:ACV}
we will encounter integrals of the following type:
\be
C_1(t)= \int_{-i\infty}^{ i \infty} \frac{d s}{2\pi i} \; A(s,t)  \; .
\label{eq:fixedpole}
\ee 
The integral (\ref{eq:fixedpole}) is defined when the amplitude
vanishes at large $s$ faster than $1/|s|$, thus satisfying an
unsubtracted dispersion relation, Eq.~(\ref{eq:DR}). Such integrals
will play a special role in our subsequent derivation of the eikonal
approximation where $A(s,t)$ is the crossing-even ``particle-Pomeron''
scattering amplitude, and this has been used extensively by Amati,
Ciafaloni and Veneziano~\cite{Amati:1987uf} in their discussion of
eikonalization for closed (super)-strings in flat space.  We will
introduce the notation of a particle-Pomeron amplitude in
Sec.~\ref{sec:eikonal} and discuss its role for eikonalization further
in Sec.~\ref{sec:twopomcut} and \ref{sec:ACV}.  Here we point out how
integral (\ref{eq:fixedpole}) arises from a $J$-plane perspective.

 Because the amplitude is crossing even, its  $s$-channel and $u$-channel discontinuities are equal, $A_s=A_u$.  The integration path runs along the imaginary $s$-axis, crossing the real axis between the $s$- and $u$-cut.  With the amplitude vanishing faster than $1/|s|$, one can distort the integration contour, e.g, one can integrate along the real axis, under the left-hand $u$-cut and over the right-hand $s$-cut. Alternatively, one can directly close the contour either to the right or to the left. When closing the contour, either to the left or to the right, one would pick up discontinuity across the respective cut,  leading to
\be
C_1(t) = (1/\pi) \int_0^\infty ds' A_s(s',t) = (1/\pi) \int_0^\infty du' A_u(u',t) 
\label{eq:fixedpoleresidue}
\ee
Historically, this contribution, $C_1(t)$,  has been referred to as the $j=-1$ ``fixed-pole'' residue.

To gain a better understanding on  this contribution, consider  the
Regge representation for the full amplitude (\ref{eq:Regge}).  As one
pushes the contour to the left in $ j$, the zero of the denominator
$\sin \pi j$  would appear to give rise to fixed powers
$s^{-N}$, $N=1, 2, \cdots$.  However for the flat space closed
string the leading term is $s^{2+ \alpha't/2}$ so these contributions
must be absent for sufficiently negative $ t$. This implies zeroes in
the numerator to cancel these poles. On the other hand we can also directly examine the amplitude with only a right-hand cut,
\be 
A_R(s,t) = \frac{1}{\pi} \int_0^\infty ds'
\frac{A_s(s',t)}{s'-s} \rightarrow \beta(t) (-\alpha' s)^{\alpha(t)} -
\frac{C_{1,s}(t) }{s} - \cdots 
\ee 
with 
$C_{1,s}(t) =   (1/\pi) \int_o^\infty ds' A_s(s',t)  \;, \label{eq:fixedpoleright}
$
and similarly 
\be 
A_L(u,t) = \frac{1}{\pi} \int_0^\infty du'
\frac{A_u(u',t)}{u'-u} \rightarrow \beta(t) (-\alpha' u)^{\alpha(t)} -
\frac{C_{1,u}(t) }{u} - \cdots 
\ee 
with 
$
C_{1,u}(t) =   (1/\pi) \int_o^\infty du' A_u(u',t)  \;.\label{eq:fixedpoleleft}
$
For closed strings, crossing symmetry $s
\leftrightarrow u$ relates the left- and right-hand
discontinuities, and it follows that $C_{1,s}(t)=C_{1,u}(t)\equiv C_1(t)$. Therefore, the fixed-pole residue is simply the coefficient of the fixed $1/s$ and $1/u$ contributions in the asymptotic expansion for $A_R$ and $A_L$ respectively. Note that the full amplitude, $A(s,t) = A_R+ A_L$, does not contain the $1/s$ term for $s$ large. However, in Eq.~(\ref{eq:fixedpole}), since the integration path runs between the left- and right-hand cuts, it cannot be distorted to infinity. As a consequence, the integral leads to  a non-vanishing  contribution, even if the full amplitude $A(s,t)$  vanishes faster than $1/|s|$.  
 
\subsection{The Pomeron in Impact Parameter Space}
\label{sec:impactspace}
The $J$-plane formalism can be applied to the Pomeron as understood in
the work of Ref.~\cite{Brower:2006ea}.  For ${\cal N} =4$ SYM, or
indeed any conformal theory dual to a string theory on $AdS_5 \times
M_5$, the Pomeron propagator, ${\cal K}(s,t,z,z')$ in
Eq.~(\ref{eq:AdSinvMellin1}), can be found at strong coupling. In
Ref.~\cite{Brower:2006ea}, we have concentrated on the imaginary part
of the full kernel, ${\rm Im}[{\cal K}]$.
In this section we transform the full kernel ${\cal K}$
to the $J$-plane, and then to transverse position space.  This leads to
remarkable simplifications.

Just as done for the flat space string theory, it is useful to
reconstruct the full amplitude through a $J$-plane representation. From
the $s$-channel discontinuity, $2i\; {\rm Im}\; {\cal K}$, one can
obtain a $J$-plane amplitude via a Mellin transform.  For the case of the
$AdS$ Pomeron, it is convenient to define the Mellin transform with respect
to 
\be
\widehat s = zz' s \; ,
\ee
where the dimensionless variable, $\widehat s$, is $R^2$ times the
proper center-of-mass energy squared.  Starting from the imaginary
part of ${\cal K}(s,t,z,z')$, obtained in \cite{Brower:2006ea}, we can
find the kernel in the $J$-plane, ${\cal K}(j,t,z,z')$, using
Eq.~(\ref{eq:MellinofImK1}).  The strong coupling kernel in the
$J$-plane, ${\cal K}(j,t,z,z')$, is a momentum space Green's function
propagating in $AdS_5$
\be 
[  -    z^5 \dd_z z^{-3} \dd_z  -  z^2 t +2\sqrt{\lambda}(j-2) ]{\cal K}(j,t,z,z')= R^{-4}z^5\delta(z-z')
\label{eq:ads5qDE}
\ee 
We can always reconstruct the full amplitude,
${\cal K}(s,t,z.z') $, using an inverse Mellin transform,
Eq.~(\ref{eq:AdSinvMellin1}).  
As in the flat-space string theory, the contour of integration
must be to the left of the poles at positive integers from the $1/\sin \pi
j$ factor, and to the right of singularities of ${\cal K}(j,t,z,z')$.

From a spectral analysis for
Eq.~(\ref{eq:ads5qDE}) in momentum space, following \cite{Brower:2006ea},
we can obtain
\be
{\cal K}(j, t,z,z')=  \frac{(zz')^2}{\pi^2R^4}\int_{-\infty}^\infty d\nu\ 
(\nu\sinh \pi \nu) 
\frac{ K_{i\nu}(qz)K_{-i\nu}(qz')}{ \nu^2  + (2\sqrt \lambda) (j- j_0)}\; ,
 \label{eq:jPomeron}
\ee
where $j_0= 2- {2}/{\sqrt \lambda}$. This 
expression masks the simplicity of the conformal
invariance, but illustrates that the $J$-plane spectrum
consists of only a continuum, with a square-root branch point at
$j_0$, i.e., ${\cal K}(j, t,z,z') \sim \sqrt {j-j_0}$, where $j_0$ is
the location of BFKL branch point in the strong coupling.  Although
the location of the BFKL cut is $t$-independent, the discontinuity
depends on $t$, for fixed $z,z'$.  Also, in the limit $t\rightarrow 0$, 
with $z,z'$ fixed, the
nature of the singularity changes: as  demonstrated in
Ref.~\cite{Brower:2006ea}, ${\cal K}(j, 0,z,z') \sim
1/\sqrt{j-j_0}$, which leads to an asymptotic behavior $\tilde
s^{j_0}/ \log^{1/2} \tilde s$, as indicated in Eq.~(\ref{eq:ImKzero}).

Now, in preparation for the eikonal application, we move
to transverse impact parameter space: $x^\perp=(x^1,x^2)$. Introducing
the conjugate transverse momentum vector, $q_\perp=(q_1,q_2)$ where $t
=- q^2_\perp$, we obtain from Eq.~(\ref{eq:jPomeron1})
\be
{\cal K}(j,x^\perp - x'^\perp ,z,z') = \int \frac{d^2q_\perp}{(2\pi)^2}  e^{i q_\perp( x^\perp - x'^\perp)} {\cal K}(j,-q_\perp^2 ,z,z') = 
\left(\frac{z z'}{R^4}\right) G_3(j,v) \; ,
\label{eq:ads3kernel}
\ee
where $G_3(j,v)$ is defined in Eq.~(\ref{eq:adschordal}).  This
elegant expression deserves an explanation, which we postpone to
Sec~\ref{sec:geo}, in order to move swiftly to the eikonal expansion.

Now using the impact-parameter-space
version of Eq.~(\ref{eq:AdSinvMellin1}), namely Eq.~(\ref{eq:pomeronkernel1}),
we can obtain ${\cal K}(s,x^\perp-x'^\perp,z,z')$,
which is the kernel in the form needed for the eikonal calculation.
We will return in Section \ref{sec:geo} to examine
 its high energy behavior more carefully.  
Here we merely remark on its phase.
Due to the BFKL branch point, 
\be
{\cal K}(s,x^\perp-x'^\perp,z,z')\  
\sim  -\frac {\widehat s^{j_0}+ (-\widehat s)^{j_0}}{\sin \pi j_0} =  -   \left(\frac {e^{-\pi j_0/2} }{\sin \pi j_0/2}\right)\;  \widehat  s^{j_0} 
\label{eq:pomeronphase}
\ee 
up to logarithmic corrections.\footnote{The expression second from the
right, though less compact than the rightmost expression, will be used
below, in order to continue to exhibit the connection between the
Regge phase and the $s$- and $u$-channel discontinuities.  
Therefore the coefficient of $s^{j_0}$ is complex, and independent of
the coordinates $b,z,z'$, in the region where the Regge form of amplitude is applicable.   Here, $(-s)^{j_0}$ and $s^{j_0}$ separately
represent asymptotic behavior for amplitudes with $s$- and $u$-channel
discontinuities.  The fact that the Pomeron kernel is complex will be
important when we discuss $s$-channel unitarity in Section
\ref{sec:unitarity}.}


\section{Eikonal Expansion of the  AdS  Pomeron}
\label{sec:eikonal}

We now turn to the problem of the eikonal summation of multiple Regge
exchange graphs for the $AdS_5$ strong coupling Pomeron. It is easy to
infer the answer by comparison with two simpler examples: the well
known eikonalization of a single Regge-pole exchange in the single
2-to-2 elastic unitarity approximation, and the recent eikonal
formula~\cite{Brower:2007qh, Cor} for graviton exchange at infinite
coupling in $AdS_5$.  The first introduces a non-trivial phase in the
Regge exchange kernel. The second brings into play the the radial
co-ordinate in $AdS_5$, which combines with the Minkowski-space impact
parameter to form an $AdS_3$ transverse space, a point we return to in
section 4.  Here and below, to simplify formulas, we temporarily set
the $AdS$ curvature radius $R$ to 1.

For pedagogical reasons we will begin by presenting the form of the eikonal
representation in $AdS_5$ before providing its derivation and a
description of its properties.  Also we will make a comparison in
Sec.~\ref{sec:ACV} with a third example, namely the eikonal
approximation for the flat space superstring amplitude,
due to Amati, Ciafaloni and
Veneziano.  Together these examples provide a general intuitive
picture to guide further advances beyond the eikonal approximation.

The standard eikonal formula takes the classic form,
\be
A(s,t) = - 2 i s \int d^2b e^{-i b^\perp q_\perp} \; \left [ e^{i\chi(s,b^\perp)} - 1\right] \; ,
\label{eq:eik}
\ee
where $t = - q^2_\perp$.  For a single Regge pole exchange, as for the
Pomeron, $\chi(s,b^\perp)$ is the Fourier transform to impact
parameter space of the elastic amplitude in the one-Reggeon exchange
approximation,
\be
\chi(s,b^\perp) = \frac{1}{2s} \int \frac{d^2q_\perp}{(2 \pi)^2} e^{i b^\perp q_\perp} A^{(1)}(s,t) \; ,
\ee
with $ A^{(1)}(s,t) = - [(  e^{- i \pi \alpha(t)}\pm 1)/\sin\pi \alpha(t)] \beta(t) s^{\alpha(t)}$.
(See also Eq.~(\ref{eq:ReggeTachyon}) for the closed string form of $A^{(1)}(s,t)$.) This is the
leading contribution to the sum of graphs depicted in Fig.~\ref{fig:eik} below.
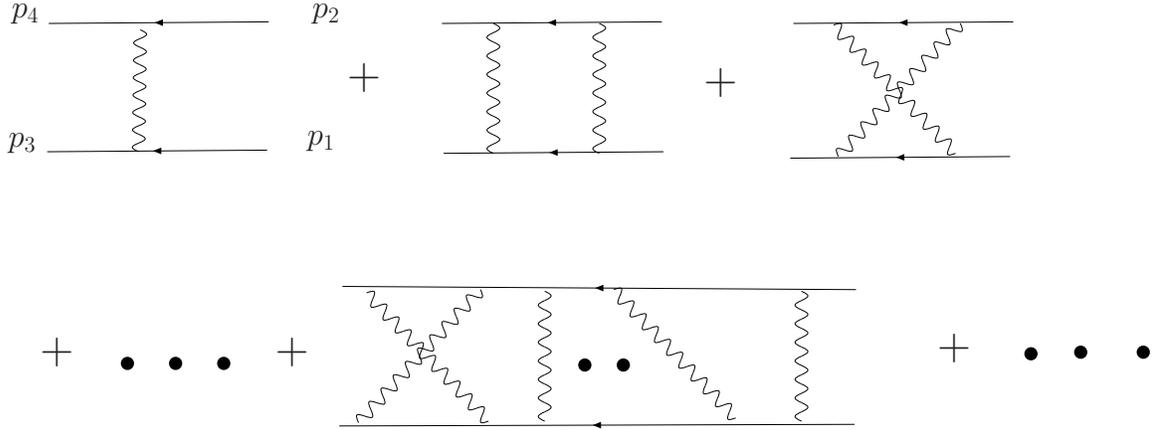
\begin{figure}[h]
\begin{center}
\scalebox{.7}{
  \begin{picture}(614,232) (18,-26)
    \SetWidth{0.5}
    \ArrowLine(558.99,191.8)(440.62,190.94)
    \ArrowLine(557.26,119.23)(438.9,118.36)
    \Photon(462.22,190.07)(527.89,120.96){3.46}{10}
    \Photon(530.48,190.94)(464.82,119.23){3.46}{10}
    \ArrowLine(369.78,191.8)(251.42,190.94)
    \ArrowLine(370.64,121.82)(252.28,120.96)
    \Photon(279.06,120.96)(279.06,190.94){3.46}{7}
    \Photon(336.08,120.96)(336.08,190.94){3.46}{7}
    \Text(393.97,152.06)[lb]{\Huge{\Black{$+$}}}
    \Text(201.31,154.65)[lb]{\Huge{\Black{$+$}}}
    \Photon(272.15,47.52)(206.49,-24.19){3.46}{10}
    \Photon(210.81,45.79)(276.47,-23.33){3.46}{10}
    \Photon(306.71,-23.33)(306.71,46.65){3.46}{7}
    \Photon(343.86,47.52)(409.52,-21.6){3.46}{10}
    \Photon(444.94,-23.33)(444.94,46.65){3.46}{7}
    \Text(520.11,8.64)[lb]{\Huge{\Black{$+$}}}
    \ArrowLine(473.46,-25.06)(196.12,-25.92)
    \ArrowLine(474.32,47.52)(197.85,49.25)
    \Vertex(328.31,6.91){3.56}
    \Vertex(349.04,6.91){3.56}
    \Text(162.43,6.05)[lb]{\Huge{\Black{$+$}}}
    \Text(35.42,6.05)[lb]{\Huge{\Black{$+$}}}
    \Vertex(82.08,7.78){3.56}
    \Vertex(108,7.78){3.56}
    \Vertex(133.92,7.78){3.56}
    \Vertex(595.28,13.82){3.56}
    \Vertex(568.49,12.96){3.56}
    \Vertex(628.97,13.82){3.56}
    \ArrowLine(158.11,191.8)(39.74,190.94)
    \ArrowLine(157.24,122.68)(38.88,121.82)
    \Photon(88.12,121.82)(88.99,187.48){3.46}{7}
    \Text(179.71,122.68)[lb]{\Large{\Black{$p_1$}}}
    \Text(182.3,191.8)[lb]{\Large{\Black{$p_2$}}}
    \Text(19.87,191.8)[lb]{\Large{\Black{$p_4$}}}
    \Text(18.14,121.82)[lb]{\Large{\Black{$p_3$}}}
  \end{picture}
}
\caption{Ladder and crossed ladder diagrams contributing to the eikonal approximation in the high    energy limit.}
\label{fig:eik}
\end{center}
\end{figure}
Let us compare this with our result for the eikonalization of the
$AdS_5$ graviton of Ref.~\cite{Brower:2007qh}
\be
\; A_{2\to2}(s,t) \simeq   - 2i s
\int d^2b \; e^{-ib^\perp q_\perp} \int dz dz' P_{13}(z) P_{24}(z') \left [ e^{i\chi(s,b^\perp, z,z')} - 1\right] 
\label{eq:adseik}
\ee
where $b = x^\perp - x'^\perp$ due to translational invariance.  The
salient new features relative to the above four-dimensional expressions
are the new transverse co-ordinate for the fifth dimension in
$AdS_5$ and the product of wave functions for right-moving ($1 \rightarrow 3)$ and
left-moving ($2 \rightarrow 4$) states,
\be 
P_{13}(z) =( z/R)^2\sqrt{g(z)}  \Phi_1(z) \Phi_3(z) \tbox{and} P_{24}(z) = (z'/R)^2\sqrt{g(z')}  \Phi_2(z') \Phi_4(z') 
\label{eq:overlapfunction}
\ee
The obvious (and correct) guess for the eikonalization of $AdS_5$ Pomeron is to simply use the
appropriate $AdS_3$ kernel for this exchange as presented above in Sec.~\ref{sec:intro},
\bel{AdSpomeronchi}
\chi(s,x^\perp- x'^\perp,z,z')= 
  \frac{ g^2_0R^4}{ 2(zz')^2 s}  {\cal K}(s,x^\perp - x'^\perp,z,z')
\ee  
where $g_0^2= \kappa_5^2/R^3$. This is a natural generalization of our
earlier result for $AdS$ graviton exchange, whose kernel can be
obtained from the same $J$-plane analysis by taking the limit $\lambda
\rightarrow \infty$, as explained in Sec.~\ref{subsec:obtaingraviton}.

\subsection{Eikonal Graphs}

In Ref.~\cite{Brower:2007qh}, we have considered the high energy limit
of a class of Witten diagrams, illustrated in Fig.~\ref{fig:eik},
where we choose scalar fields for the external lines along two sides
of the ladder and gravitons for the exchanged rungs between these two
sides.  The sum includes all possible $AdS$ graviton exchanges,
crossed and uncrossed.  The treatment of the eikonal sum for conformal
Pomeron exchanges follows exactly as that for the $AdS$ graviton. The
only new ingredient is to replace each $AdS$ graviton propagator by a
conformal Pomeron propagator, ${\cal K}(s,x^\perp-x'^\perp, z,z')$,
Eq.~(\ref{eq:pomeronkernel1}). Because we work in to leading order in
strong coupling, we can again treat the two scattered particles ---
the sides of the ladder --- by using an $AdS_5$ scalar propagator, as
was done in Ref.~\cite{Brower:2007qh}.  String excitations on the
sides of the ladder enter at higher order in $1/\sqrt \lambda$, and
can be ignored for our current purposes.

Most of the needed analysis was done in
Ref.~\cite{Brower:2007qh} and will not be repeated
here. We only outline briefly how the eikonal sum
can be carried out, though  we will spell out explicitly the Feynman
rules for the eikonal graphs. For a 2-to-2 amplitude,
$1+2\rightarrow 3+4$, let us denote the longitudinal momenta by
$p_1^{\pm} + p_2^{\pm} \rightarrow p_3^{\pm} + p_4^{\pm}$, with an 
all-incoming convention.  We will work in a transverse coordinate
basis, using $p^+,p^-, x^\perp,z$ as coordinates. In this representation,
after stripping away a wave function for each external particle,
$\Phi_i(z)e^{-ip^\perp x^\perp}$, we will be left to calculate an
amputated amplitude, ${\cal A}(p_i^\pm, x^\perp_i,z_i)$, as a
perturbative sum of diagrams illustrated in Fig.~\ref{fig:eik}.

In the high-energy near-forward limit, $p_1^+\simeq -p^+_3$ and
$p_2^-\simeq -p^-_4$ are large, with $q^\pm= p^\pm_1+p^\pm_3= -
(p^\pm_2+p^\pm_4) = 0(1/\sqrt{p^+_1p^-_2})$.  Therefore, ${\cal A}$
depends on longitudinal momenta only through $s\simeq 2p_1^+p_2^-$,
and we can simply express the amplitude as ${\cal A}(s, x^\perp_i,
z_i)$. It is in fact useful to view this as matrix elements of an
operator, $\bf A$, in transverse coordinate basis, 
\be {\cal A}(s,
x^\perp_i, z_i)= \< x_3,z_3, x_4,z_4| {\bf A}| x_1,z_1, x_2,z_2 \>=\<3,4|{\bf A}| 1,2 \> \; ,
 \ee
with states normalized by
\be
\< 3,4| 1,2 \>= [ \delta^2(x^\perp_1-x^\perp_3)
\delta(z_1-z_3)/\sqrt g_1][\delta^2(x^\perp_2-x^\perp_4) \delta(z_2-z_4)/\sqrt g_2] \; .
\ee
Perturbative diagrams can be organized by the number of Pomeron propagators exchanged.  

Let us begin by examining the simplest diagram -- the tree
graph. Since this involves a single Pomeron propagator, the amputated
amplitude in transverse coordinate representation is given by
\be
{\cal A}^{(1)}(s, x_i^\perp,z_i)
= 2 \left(\frac{z_1z_2}{R^2}\right)^2 s \; \chi (s,x^\perp_1-x^\perp_2, z_1,z_2)\<3,4| 1,2\> \; .
\ee
The $\<3,4|1,2\>$ factor is supplied so as to reproduce the single
Pomeron exchange contribution obtained in Sec.~\ref{sec:regge}. Note
that ${\cal A}^{(1)}$ is diagonal in the transverse coordinate basis.

\subsection{One-Loop Contribution:}

\begin{figure}
\begin{center}


\scalebox{0.65}{
  \begin{picture}(614,318) (9,-12)
    \SetWidth{1.0}
    \ArrowLine(307.59,257.03)(283.89,280.73)
    \ArrowLine(290.73,185.4)(312.33,208.04)
    \SetWidth{2.0}
    \Line(313.38,209.62)(313.33,254.39)
    \SetWidth{0.5}
    \Vertex(313.91,208.04){3.83}
    \Vertex(313.38,255.97){3.83}
    \Vertex(313.38,255.97){3.83}
    \Vertex(313.38,255.97){3.83}
    \Vertex(312.86,255.97){3.83}
    \Vertex(312.86,255.97){3.83}
    \Vertex(312.86,255.97){3.83}
    \SetWidth{1.0}
    \SetWidth{2.0}
    \Line(388.17,208.04)(388.12,252.81)
    \SetWidth{0.5}
    \Vertex(386.59,207.52){3.83}
    \Vertex(388.17,255.45){4.25}
    \SetWidth{1.0}
    \Photon(316.02,257.03)(384.49,209.1){2.9}{9}
    \Photon(316.54,206.46)(385.01,253.87){2.9}{10}
    \ArrowLine(392.91,260.19)(419.77,286.52)
    \ArrowLine(69,260.71)(45.3,284.41)
    \ArrowLine(52.14,189.08)(73.74,211.73)
    \SetWidth{0.5}
    \Vertex(75.32,211.73){3.83}
    \Vertex(74.79,259.66){3.83}
    \Vertex(75.32,211.73){3.83}
    \Vertex(74.79,259.66){3.83}
    \Vertex(75.32,211.73){3.83}
    \Vertex(74.79,259.66){3.83}
    \Vertex(74.79,211.73){3.83}
    \Vertex(74.26,259.66){3.83}
    \Vertex(74.79,211.73){3.83}
    \Vertex(74.26,259.66){3.83}
    \Vertex(74.79,211.73){3.83}
    \Vertex(74.26,259.66){3.83}
    \SetWidth{1.0}
    \Photon(77.42,260.71)(149.05,258.61){2.9}{8}
    \Photon(77.95,210.15)(143.26,209.62){2.9}{8}
    \SetWidth{2.0}
    \Line(149.58,211.73)(149.53,256.5)
    \SetWidth{0.5}
    \Vertex(148,211.2){3.83}
    \Vertex(149.58,259.13){3.83}
    \SetWidth{1.0}
    \ArrowLine(150.11,264.4)(176.97,290.73)
    \Text(26.86,171.17)[lb]{\Large{\Black{$p_1$}}}
    \SetWidth{2.0}
     \Line(73.74,211.2)(73.68,255.97)
    \SetWidth{1.0}
    \ArrowLine(174.34,184.87)(150.63,208.57)
    \ArrowLine(413.45,180.66)(389.75,204.36)
    \Text(24.23,288.63)[lb]{\Large{\Black{$p_3$}}}
    \Text(185.4,174.86)[lb]{\Large{\Black{$p_2$}}}
    \Text(187.5,297.58)[lb]{\Large{\Black{$p_4$}}}
    \Text(273.35,288.63)[lb]{\Large{\Black{$p_3$}}}
    \Text(274.93,171.17)[lb]{\Large{\Black{$p_1$}}}
    \Text(426.62,172.76)[lb]{\Large{\Black{$p_2$}}}
    \Text(431.36,292.31)[lb]{\Large{\Black{$p_4$}}}
    \Text(231.22,234.9)[lb]{\Large{\Black{$+$}}}
    \Text(489.3,239.12)[lb]{\Large{\Black{$=$}}}
    \SetWidth{2.0}
    \Line(77.42,69)(76.9,24.23)
    \SetWidth{1.0}
    \ArrowLine(53.2,0)(74.79,22.12)
    \SetWidth{0.5}
    \Vertex(76.9,21.07){4.5}
    \SetWidth{1.0}
    \ArrowLine(76.9,79)(53.2,102.7)
    \SetWidth{0.5}
    \Vertex(76.9,74.26){4.5}
    \SetWidth{1.0}
    \ArrowLine(204.36,79)(180.66,102.7)
    \SetWidth{0.5}
    \Vertex(205.94,79){4.5}
    \SetWidth{1.0}
    \ArrowLine(178.55,1.58)(200.14,23.7)
    \SetWidth{0.5}
    \Vertex(203.83,27.39){4.5}
    \SetWidth{1.0}
    \SetWidth{2.0}
    \Line(204.36,76.37)(204.88,30.02)
    \SetWidth{1.0}
    \Photon(80.58,21.07)(115.35,22.12){2.63}{4}
    \Photon(78.48,79)(110.61,78.48){2.63}{3}
    \Photon(210.15,80.06)(247.02,23.7){2.63}{7}
    \Photon(209.1,23.7)(244.39,82.69){2.63}{8}
    \Text(119.03,84.27)[lb]{\Large{\Black{$q_2$}}}
    \Text(123.77,13.69)[lb]{\Large{\Black{$q_1$}}}
    \Text(30.02,-12.64)[lb]{\Large{\Black{$p_1$}}}
    \Text(160.11,-11.59)[lb]{\Large{\Black{$p_1$}}}
    \Text(251.76,81.64)[lb]{\Large{\Black{$q_2$}}}
    \Text(258.08,18.96)[lb]{\Large{\Black{$q_1$}}}
    \SetWidth{0.5}
    \qbezier(22.12,-5.27),(11.06,52.67),(26.86,103.23)
    \qbezier(278.09,2.11),(291.79,57.41),(273.88,107.97)
    \Text(139.57,49.51)[lb]{\Large{\Black{$+$}}}
    \Text(169.07,107.97)[lb]{\Large{\Black{$p_3$}}}
    \Text(34.76,107.97)[lb]{\Large{\Black{$p_3$}}}
    \Text(0.6,40.83)[lb]{\Large{\Black{$\frac{1}{2}$}}}
    \Text(327.6,51.83)[lb]{\Large{\Black{$\times$}}}
    \Vertex(551.97,30.55){4.5}
    \Vertex(553.55,79.53){4.5}
    \SetWidth{1.0}
    \SetWidth{2.0}
    \Line(554.03,78.48)(554.08,33.18)
    \SetWidth{1.0}
    \Photon(549.34,76.9)(508.78,30.02){2.63}{6}
    \Photon(545.65,32.13)(507.2,82.69){2.63}{7}
    \ArrowLine(553.55,82.16)(577.25,109.03)
    \ArrowLine(577.25,1.05)(554.61,25.28)
    \Text(466.12,54.78)[lb]{\Large{\Black{$+$}}}
    \Text(368.68,81.64)[lb]{\Large{\Black{$q_2$}}}
    \Text(485.08,82.16)[lb]{\Large{\Black{$q_2$}}}
    \Text(372.9,29.49)[lb]{\Large{\Black{$q_1$}}}
    \Text(485.08,25.28)[lb]{\Large{\Black{$q_1$}}}
    \Text(462.96,-4.74)[lb]{\Large{\Black{$p_2$}}}
    \Text(464.54,117.45)[lb]{\Large{\Black{$p_4$}}}
    \Text(580.41,114.82)[lb]{\Large{\Black{$p_4$}}}
    \Text(585.16,-6.85)[lb]{\Large{\Black{$p_2$}}}
    \SetWidth{2.0}
    \Line(429.25,83.74)(429.25,38.98)
    \SetWidth{0.5}
    \Vertex(429.25,35.29){4.5}
    \Vertex(427.15,82.69){4.5}
    \SetWidth{1.0}
    \Photon(423.99,82.16)(395.02,80.58){2.63}{3}
    \Photon(422.93,35.82)(401.34,35.29){2.63}{2}
    \ArrowLine(430.83,86.9)(455.59,112.19)
    \ArrowLine(455.06,2.11)(431.89,30.02)
    \SetWidth{0.5}
\qbezier(614.12,1.05),(622.55,56.88),(604.64,113.24)
\qbezier(370.27,5.79),(351.84,55.83),(368.17,107.97)
  \end{picture}
}

\caption{Sum of box and crossed box diagrams is factorized with combinatoric
weight $1/2!$.}
\label{fig:box}
\end{center}
\end{figure}
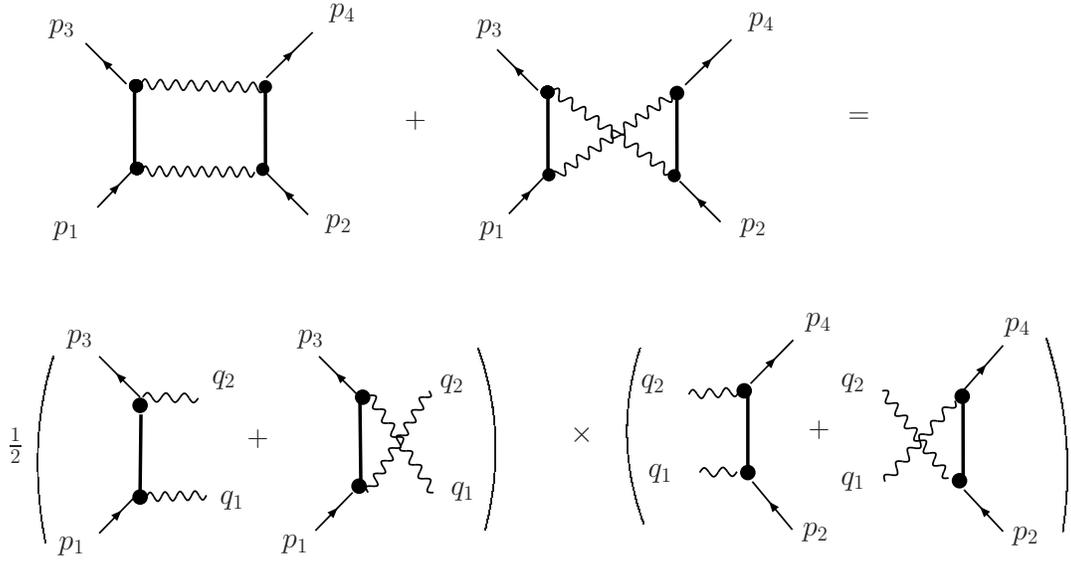

\begin{figure}
\begin{center}
\scalebox{0.50}{
  \begin{picture}(614,196) (29,-34)
    \SetWidth{0.5}
    \Text(396.93,152.42)[lb]{\Large{\Black{$q_2$}}}
    \Text(614.12,149.24)[lb]{\Large{\Black{$q_2$}}}
    \Text(465.52,151.15)[lb]{\Large{\Black{$p_3$}}}
    \SetWidth{0.5}
    \GOval(105.42,67.95)(44.46,20.96)(0){0.882}
    \ArrowLine(58.43,-6.35)(92.72,30.48)
    \ArrowLine(58.43,-6.35)(92.72,30.48)
    \ArrowLine(58.43,-6.35)(92.72,30.48)
    \ArrowLine(58.43,-6.35)(92.72,30.48)
    \SetWidth{1.0}
    \ArrowLine(59.06,-6.35)(93.36,30.48)
    \SetWidth{0.5}
    \Photon(116.22,29.85)(154.33,-8.26){4.76}{4}
    \SetWidth{1.0}
    \Photon(116.22,29.85)(154.33,-8.26){4.76}{4}
    \Photon(163.85,144.16)(116.22,106.06){4.76}{4}
    \ArrowLine(97.17,106.06)(59.06,144.16)
    \Text(182.9,144.16)[lb]{\Large{\Black{$q_2$}}}
    \ArrowLine(97.17,106.06)(59.06,144.16)
    \Text(267.37,144.8)[lb]{\Large{\Black{$p_3$}}}
    \ArrowLine(59.06,-6.99)(93.36,29.85)
    \ArrowLine(302.3,-10.8)(336.59,26.04)
    \Photon(391.21,143.53)(343.58,105.42){4.76}{4}
    \Photon(344.85,24.13)(382.95,-13.97){4.76}{4}
    \Photon(526.48,31.12)(599.52,150.51){4.76}{9}
    \Text(625.56,-27.31)[lb]{\Large{\Black{$q_1$}}}
    \Text(29.85,146.07)[lb]{\Large{\Black{$p_3$}}}
    \Text(266.73,-32.39)[lb]{\Large{\Black{$p_1$}}}
    \Text(30.48,-24.13)[lb]{\Large{\Black{$p_1$}}}
    \Text(173.38,-24.13)[lb]{\Large{\Black{$q_1$}}}
    \Text(392.48,-34.29)[lb]{\Large{\Black{$q_1$}}}
    \Text(463.61,-34.29)[lb]{\Large{\Black{$p_1$}}}
    \Text(229.26,62.24)[lb]{\Large{\Black{$\simeq$}}}
    \Text(431.86,64.14)[lb]{\Large{\Black{$+$}}}
    \SetWidth{0.5}
    \Vertex(339.77,26.04){6.02}
    \Vertex(520.13,27.31){6.02}
    \Vertex(520.13,104.15){6.02}
    \Vertex(338.5,103.52){6.02}
    \SetWidth{3.0}
    \Line(338.5,102.88)(338.5,31.12)
    \Line(519.5,103.52)(519.5,31.75)
    \SetWidth{1.0}
    \Photon(527.12,104.15)(616.67,-12.07){4.76}{11}
    \ArrowLine(332.15,107.33)(294.04,145.43)
    \ArrowLine(518.23,107.33)(480.12,145.43)
    \ArrowLine(483.93,-12.07)(518.23,24.77)
  \end{picture}
}
\caption{Reggeon particle scattering amplitude in the planar approximation, $A_{13}(M^2,t)$, with $t = (q_1 + q_2)^2$ and $M^2 = (p_1 + q_1)^2$.}
\label{fig:pompart}
\end{center}
\end{figure}
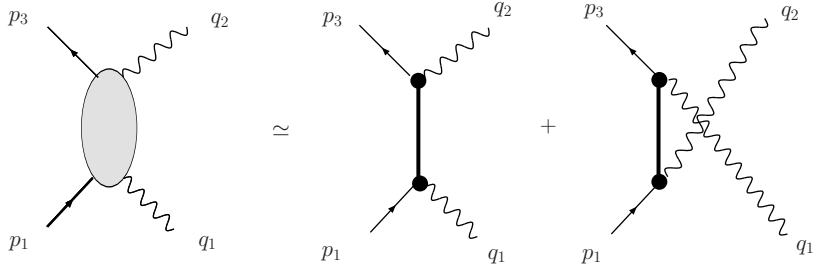

Before presenting the result for general graphs with $n$ Pomeron
exchanges, we first consider the one-loop contribution, illustrated in
Fig.~\ref{fig:box}. There are two independent diagrams, which are depicted in the upper
half of Fig.~\ref{fig:box}. For reason to be clarified shortly, 
the sum of these two diagrams can be
combined as a product of two  ``Pomeron-particle'' amplitudes, 
$A_{13}$ and $A_{24}$, (see Fig.~\ref{fig:pompart}),
connecting through two Pomeron kernels, as schematically represented
by the lower half of Fig.~\ref{fig:box}.  However, this leads to an
over-counting, and a factor of $1/2!$ is supplied.

 It is important to appreciate the assumptions
made in evaluating the one-loop contribution in the high energy eikonal
approximation.  We assume that it is proper to factorize the
contribution into exchange Pomeron kernels  for the rungs of
the ladder and 2-2 Pomeron-particle scattering amplitudes $A_{13}$ and 
$A_{24}$ on the sides.  In an elementary field theory, 
e.g., the eikonal sum for exchanging conformal gravitons, this is a trivial
combinatoric fact as illustrated in Fig.~\ref{fig:box}.  In string theory
this is an assumption on the high energy limit of the one loop
diagram, only proven for the flat space superstring to
date~\cite{Amati:1987uf}. However we should note that the existence of
these Pomeron-particle amplitudes is supported by the observation in
Ref.~\cite{Brower:2006ea} that the Pomeron vertex is a proper on-shell
vertex operator with conformal weight (1,1) in string theory, both in flat
space and to leading order in an $AdS$ background.

The Pomeron-particle elastic-scattering
amplitude in the planar limit can be expressed through a dispersion
relation as a sum of $s$-channel and $u$-channel closed string exchanges (see
Fig.~\ref{fig:pompart}).
As we noted, at the leading order in strong coupling, we can represent 
this amplitude using the scalar propagator,
$G_5(2p^+p^-,x^\perp-x'^\perp, z,z')$, in the bulk of $AdS_5$.  This
scalar propagator is the solution of Eq.~(\ref{eq:ads5qDE}) at $j=2$,
but normalized without a factor of $R^{-4}$, so it has the scaling
dimension of $({\rm length})^2$. Again we use a transverse coordinate
representation, expressing $G_5$ in terms of $p^\pm$ and
$x^\perp-x'^\perp$.  A useful spectral representation for $G_5$ is
\be
G_5(2p^+p^-, x^\perp-x'^\perp,z,z')=\frac{(zz')^2}{2}\int \frac {d^2 p^\perp }{ (2\pi)^2}
e^{ip^\perp (x^\perp-x'^\perp)} \int_0^\infty  dk^2 \frac {J_2(zk) J_2(z'k)}{ k^2 +{ p^\perp}^2 -2p^+p^- -i\epsilon } \label{eq:G5mixedrep} \ .
\ee
Following a by-now standard procedure~\cite{Amati:1987uf,CW,Chang:1971je}, 
the total one-loop contribution at high energies can be expressed as,
\bea
{\cal  A}^{(2)} ( p^\pm_i,x_i^\perp,z_i) &=& - i \frac{ i^4 g_0^4 }{2!} \int dq_1 dq_2 
A_{13}(p_1^\pm, q_1^\pm, x^\perp_3-x^\perp_1, z_3, z_1)   
 \; {\cal K}(s,x^\perp_1-x^\perp_2, z_1, z_2) \nn 
&&\; {\cal K}(s,x^\perp_3-x^\perp_4, z_3, z_4)\;  A_{24}((p_2^\pm, q_2^\pm, x^\perp_4-x^\perp_2, z_4, z_2 ) 
\label{eq:one-loop}
\eea
where the phase space is written symmetrically as
\be
\int dq_1 dq_2 \equiv \int \frac{dq^+_1 dq^-_1}{2 \pi} \frac{dq^+_2 dq^-_2}{2 \pi} \; \delta(q^+_1+q^+_2-q^+)  \delta(q^-_1+q^-_2-q^-) 
\ee
with $q^\pm_1$ and $q^\pm_2$ the longitudinal momentum associated with
the two Pomeron exchanges.

As emphasized earlier, in the near forward limit, we have $q^\pm\simeq
0$, so in fact, $q_1^\pm=-q_2^\pm$.  The structure for ${\cal
  A}^{(2)}$ is identical to that for the exchange of two $AdS$
gravitons in Ref.~\cite{Brower:2007qh}, with conformal Pomeron
propagators replacing $AdS$ graviton propagators. As noted earlier,
the Pomeron propagators, ${\cal K}$, are independent of longitudinal
momenta, $q_i^\pm$ and can be taken outside of the integrals.

In Eq.~(\ref{eq:one-loop}), $A_{13} $ and $A_{24}$ are
Pomeron-particle amplitudes mentioned earlier and, in strong
coupling, each reduces to a sum of two scalar propagators $G_5$,
reflecting direct and crossed exchanges along each side of the ladder,
e.g.,
\be
A_{13}(p_1^\pm, q_1^\pm, x^\perp_3-x^\perp_1, z_3, z_1)  = \frac{1}{R^3}\left[G_5(p_1^\pm, q_1^\pm, x^\perp_3-x^\perp_1, z_3, z_1)  +G_5(p_1^\pm, q_2^\pm, x^\perp_3-x^\perp_1, z_3, z_1)\right]  
\ee
At high energy,  ($p_1^+$
and $p_2^-$ large with $p_1^-\sim 0$ and $p_2^+\sim 0$),  $A_{13}$ 
depends only on the integration variable $q_1^-$ through the combination $p_1^+q_1^-$ and $A_{24}$ only on
$q_1^+$ through $p_2^-q_1^+$, so that  integrals $\int dq_1^- A_{13}$ and $\int q_1^+ A_{24}$ 
can be carried out independently.  This ``left-right'' factorization is one of the key properties which
allows eikonalization to proceed. From the spectral representation for $G_5$, and the completeness relation, one arrives at a remarkably simple result, 
\be
\left(\int \frac{dq_1^- }{2 \pi i}\; A_{13}\right)\left( \int \frac{dq_1^+}{2 \pi i} \; A_{24}\right) 
=  (1/2sR^6) \delta^2 (x_1^\perp- {x_3}^\perp) z_1^3 \delta(z_1-z_3)\delta^2 (x_2^\perp- {x_4}^\perp) z_2^3 \delta(z_2-z_4)\\
\ee
Putting all terms together, we obtain
\bea
{\cal A}^{(2)}(s, x_i^\perp,z_i)&=& <3,4|{\cal A}^{(2)}| 1,2>\nonumber\\
&\simeq&- 2 i (zz'/R^2)^2\; s \; \frac{1}{ 2!} \; \left[  i \chi(s,x^\perp-x'^\perp,z,z') \right]^2<3,4| 1,2>  \nonumber\\
\label{eq:eikonalcontraction2}
\eea
This has been discussed carefully in Ref.~\cite{Brower:2007qh},
to which the reader is referred for more details.

\subsection{Feynman Rules  and Eikonal Sum:}

Feynman rules for higher order eikonal graphs can be written down
fairly simply.  In $n$th order, there are $n$ $AdS_3$ vertices on each
side to the ladder. Each vertex on one side is connected to one and
only one vertex on the opposite side by a conformal Pomeron
propagator; there are $n!$ such distinct graphs. The Feynman rules
are:
\begin{enumerate}
\item A Pomeron kernel, $ {\cal K}(s,x^\perp-x'^\perp, z,z') $, for each rung across the 
ladder,
\item An $AdS_5$ scalar propagator, $ R^{-3} G_5( 2p^+p^-, x^\perp_{i+1}- x_i^\perp,z_{i+1} ,z_i)$, 
connecting  each adjacent vertices  along the side of the ladder.
\item A factor of $g_0$ for each vertex, and a factor of $i$ for each propagator.
\item An overall  factor of $-i $.
\end{enumerate}
To calculate the $n$th order amputated amplitude, ${\cal
  A}^{(n)}(s,x_i^\perp,z_i)$, one integrates over all loop
longitudinal momenta, with momentum conservation at each vertex.  One
also integrates all internal transverse coordinates, except for
$(x^\perp_i, z_i)$, $i=1,2,3,4$, to which external momentum $p_i^\pm$
are attached. Summing over $n$ leads to the total amputated amplitude
${\cal A}(s, x^\perp_i, z_i)$.  To obtain the momentum space scattering
amplitude, $A(s,t)$ from ${\cal A}(s, x^\perp_i, z_i)$,  one supplies
external wave functions, $ e^{-iq^\perp_i x^\perp_i}\Phi(z_i)$,  and integrates over the
transverse coordinates.

It suffices to point out   that the evaluation for  higher order contributions proceeds also as has been 
done for $AdS$ graviton exchange. As stressed in Ref.~\cite{Brower:2007qh}, for each order in a perturbative eikonal sum, 
one can  again demonstrate the feature of ``zero transverse deflection,'' and the amplitudes 
becomes diagonal in the transverse basis. To be precise, we find that the amputated amplitude, at each order, takes on the form
\bea
{\cal A}^{(n)}(s, x_i^\perp,z_i)&=& <3,4|{\cal A}^{(n)}| 1,2>\nonumber\\
&\simeq&- 2 i (zz')^2\; s \; \frac{1}{ n!} \; \left[  i \chi(s,x^\perp-x'^\perp,z,z') \right]^n<3,4| 1,2>  \nonumber\\
\label{eq:eikonalcontraction}
\eea
Summing over $n=1,2,\cdots$, leads to the desired eikonal
representation. After integrating out $(x^\perp_i,z_i)$,
$i=1,\cdot\cdot,4,$ and removing a two-dimensional delta-function
associated with the center-of-mass motion in impact space, we arrive
at the eikonal representation, Eq.~(\ref{eq:adseik}), with the eikonal
given by Eq.~(\ref{AdSpomeronchi}), as promised.


\section{Conformal geometry at High Energies}
\label{sec:geo}

We now turn to a more detailed consideration of the Pomeron kernel
with an emphasis on the consequences of conformal invariance for high
energy amplitudes. This not only explains the simple properties of the
kernel, it also gives a geometrical picture of
Regge scattering in $AdS$ space.

To see intuitively how this picture comes about let us reconsider the
Regge limit for a general $n$-particle scattering amplitude:
$A(p_1,p_2, \cdots p_n)$.  As argued in Ref.~\cite{Brower:2006ea}, the
general Regge exchange corresponds to a large rapidity gap separating
the $n$ particles into two sets: the right movers and left movers,
with large $p^+_r = (p^0_r+ p^3_r)/\sqrt{2}$ and large $p^-_\ell = (p^0_\ell
- p^3_\ell)/\sqrt{2}$ momenta respectively.  The Pomeron exchange kernel
is obtained by applying this limit to the leading diagram, in the
$1/N$ expansion, that carries vacuum quantum numbers in the
$t$-channel. The Pomeron exchange graph in string theory is a
cylinder, a $t$-channel closed string.

The rapidity gaps, $\ln(p^+_r p^-_\ell)$, between any right- and
left-moving particles are all $O(\log s)$, i.e., a large Lorentz
boost, $\exp[y M_{+-}]$, with $y\sim \log s$, is required to switch
from the frame comoving with the left movers to the frame comoving
with the right movers. The $J$-plane is conjugate to rapidity, and as
such is identified with the eigenvalue of the Lorentz boost generator
$M_{+-}$. In the context of the AdS/CFT correspondence, it is
illuminating to consider the boost operator relative to the full
$O(4,2)$ conformal group, which are represented as isometries of
$AdS_5$.

The conformal group O(4,2) has 15 generators:
$P_\mu,M_{\mu\nu}, D, K_\mu$.  In terms of transformations on
light-cone variables, there are two interesting 6 parameter
subgroups: The first is the well known collinear group $SL_L(2,R)
\times SL_R(2,R)$ used in DGLAP,  with generators
\be 
SL_L(2,R), \; SL_R(2,R) \quad 
\mbox{generators:}\quad  D \pm M_{+-} \; , \; P_\pm \; , \; K_\mp  \; ,  
\ee
which corresponds in the dual $AdS_5$ bulk 
to isometries of the Minkowski $AdS_3$ light-cone sub-manifold. The second is $SL(2,C)$ (or M\"obius invariance used in solving the
weak coupling BFKL equations)  with generators
\be
SL(2,C)  \quad \mbox{generators:} \quad  i D \pm M_{12} \; ,\;  
P_1 \pm  i P_2 \; , \;  K_1 \mp i K_2 \;,
\ee
corresponding to the isometries of the Euclidean (transverse) $AdS_3$
subspace of $AdS_5$; Euclidean $AdS_3$ is the hyperbolic space $H_3$.
Indeed $SL(2,C)$ is the subgroup generated by all elements of the
conformal group that commute with the boost operator, $M_{+-}$ and as
such plays the same role as the little group which commutes with the
energy operator $P_0$. For example we note that the BFKL Pomeron
kernel in the $J$-plane is a solution of an $SL(2,C)$ invariant
operator equation in both strong and weak coupling. Very likely this
is a generic property for the Pomeron in all conformal gauge theories.

\subsection{SL(2,C) Invariance of Pomeron kernel}

In high energy small-angle scattering, the problem separates into the
longitudinal and transverse directions relative to the momentum
direction of the incoming particles.  The transverse subspace of
$AdS_5$ is $AdS_3$.  We shall show that this is reflected in
the co-ordinate representation for the Pomeron kernel of
Ref.~\cite{Brower:2006ea} as a bulk-to-bulk scalar propagator in the
transverse Euclidean $AdS_3$ with $SL(2,C)$ isometries.

Recall that our conformal strong coupling Pomeron kernel
(or ``Reggized $AdS_5$ graviton'') from Ref.~\cite{Brower:2006ea}
was written  in momentum space as an $AdS_5$ Green's function,
\be 
[  -    z^5 \dd_z z^{-3} \dd_z  -  z^2 t + 2\sqrt{\lambda}(j-2) ]{\cal K}(j,t,z,z')= R^{-4} z^5\delta(z-z') \; .
\label{eq:ads5qDE1a}
\ee 
with $AdS_5$ mass squared $2\sqrt{\lambda}(j-2)/R^2$. However in practice
one can use an impact parameter representation in which the Pomeron
kernel can be re-expressed in terms of an $AdS_3$ Green's function,
\be
{\cal K}(j,x^\perp - x'^\perp ,z,z') = 
\left(\frac{z z'}{R^4}\right) G_3(j,v) \; ,
\ee
as noted in the Introduction. Here $G_3(j,v)$ has a simple closed-form
expression (\ref{eq:adschordal}) as a function of the $AdS_3$ chordal
distance, $v = [(x_\perp-x'_\perp)^2+(z-z')^2]/2 zz'$, that
greatly simplified our subsequent analysis of multi-Pomeron
exchange.  Let us explain this ``accident'' in more geometrical
terms.

The singularities in the $J$-plane must be determined by the eigenvalues
of the boost operator, which for our $AdS$ Pomeron\footnote{In
  Ref.~\cite{Brower:2006ea} the eigenvalue condition $M_{+-}=j$ was
  also identified with the on-shell condition for the world sheet
  dilatation: $L_0 + \bar L_0 - 2=0$. Here we are concerned with the
  target space isometries.}  is approximated by $M_{+-} = 2 -
H_{+-}/(2 \sqrt{\lambda}) + O(1/\lambda)$ to leading order in strong
coupling.  In this context we find that the $AdS_3$ Green's
function for the Pomeron obeys the differential equation,
\be
[  H_{+-}  +2\sqrt{\lambda}(j-2) ]G_3(j,x_\perp-x'_\perp, 
z,z')=z^3\delta(z-z')\delta^2(x_\perp-x'_\perp)
\label{eq:ads3xDE}
\ee
for the boost operator, where
\be
H_{+-} = -z^3\dd_z z^{-1} \dd_z   - z^2  \nabla^2_{x_\perp} +3 \ .
\ee
To relate this to our earlier expression~\cite{Brower:2006ea}
for the Pomeron co-ordinate space kernel as an $AdS_5$ Green's function,  
\be
[  -   z^5 \dd_z z^{-3} \dd_z  + z^2(\dd_+ \dd_- - \nabla^2_{\perp}) 
+2\sqrt{\lambda}(j-2) ]G_5(j,x-x', z,z')= z^5\delta(z-z')\delta^4(x-x') \; ,
\label{eq:ads5xDE}
\ee
we need to recognize that in the Regge limit this exchange kernel couples
exclusively to nearly light-like left- and right-moving
sources. Consequently to compute the high energy amplitude, we 
only need the $AdS_5$ kernel projected onto these sources,
\be
 \int \frac{d x^+dx^-}{zz'} G_5(j,x-x', z,z') = G_3(j,x_\perp - 
x'_\perp, z,z') \; ,
\label{eq:G3rep}
\ee
which is precisely the $AdS_3$ kernel as can be readily seen by
integrating Eq.~(\ref{eq:ads5xDE}).  The integration measure, $dx_+
dx_-/(zz')$, ensures the result is both Lorentz boost and scale
invariant. Equivalently this approximation can be stated as
restricting the exchanged momentum to the transverse plane, so that $t
= - q^2_\perp$.  Then the $AdS_5$ momentum-space equation of motion
for Pomeron kernel, Eq.~(\ref{eq:ads5qDE}), becomes
\be
[- z^5 \dd_z z^{-3} \dd_z + z^2 q^2_\perp + 2 \sqrt{\lambda}(j-2)] {\cal 
K}(j,t,z,z')
 = (z^5/R^4) \delta(z-z')  \; .
\ee
Merely setting $q^{\pm} = 0$ reduces this to the $AdS_3$ transverse
momentum space $AdS_3$ kernel, $G_3(j,q_\perp,z,z') = (R^4/zz'){\cal K}(j,t,z,z')$, as it must. 

In order to gain a better understanding on the emergence of the $AdS_3$
algebra, let us discuss the symmetry of the scalar propagator, $G_3(j,v)$, 
in terms of the Euclidean $AdS_3$ metric, 
\be
ds^2 = \frac{R^2}{z^2} [ dz^2 + dx_1 dx_1 + dx_2 dx_2  ] = ds^2 = 
\frac{R^2}{z^2} [ dz^2 + dw d\bar w] \; ,
\ee
where the transverse subspace is $(w = x_1 +i x_2,z)$. The
generators of the $SL(2,C)$ isometries of $AdS_3$ are
\bea
J_0 &=& w \dd_w  + \Half  z\dd_z  \quad, \quad J_- = - \dd_w  \quad, 
\quad J_+ = w^2 \dd_w  + w z\dd_z - z^2 \dd_{\bar w} \nn
\bar J_0 &=& {\bar w} \dd_{\bar w}  + \Half z\dd_{z}  \quad, \quad \bar  
J_- = - \dd_{\bar w}  \quad, \quad  \bar J_+ = {\bar w}^2 \dd_{\bar w}  
+  {\bar w}  z\dd_{z} -  z^2\dd_w  \label{eq:sl2c} \; .
\eea
In the conformal group, this corresponds to the identification,
\bea
 J_0\; , \; J_+\; , \; J_-  &\leftrightarrow&  ( - i D + M_{12})/2\; , \;(P_1 + i P_2)/2\; , \;(K_1 -iK_2)/2 \nn
 \bar J_0\; , \; \bar J_+\; , \; \bar J_-  &\leftrightarrow&  (- i D - M_{12} )/2\; , \;(P_1 - i P_2)/2\; , \;(K_1 + iK_2)/2 \nonumber \; ,
\eea
so that the non-zero commutators in $SL(2,C)$  must be
$[J_0,J_\pm] = \pm J_\pm$, $[J_+, J_-] = 2 J_0$, and $[\bar J_0, \bar J_\pm]
= \pm \bar J_\pm$, $[\bar J_+, \bar J_-] = 2 \bar J_0$.  In general,
unitary representations of $SL(2,C)$ are labeled by $h = i
\nu + (1+n)/2 $, and $\bar h = i \nu +(1-n)/2$, which are the eigenvalues 
for the highest-weight state of $J_0$ and $\bar
J_0$.  The principal series is given by real $\nu$ and integer $n$.
The quadratic Casimirs $J^2$ and $\bar J^2$ 
have eigenvalues $h(h-1)$
and $\bar h (\bar h -1)$ respectively. In the representation
(\ref{eq:sl2c}) they are 
\be
J^2 = J^2_0 + \Half (J_+J_- + J_- J_+) = \frac{1}{4}[z^2 \dd^2_z  -   z 
\dd_z   + 4z^2 \dd_w\dd_{\bar w}]
\ee
with $\bar J^2 = \bar J^2_0 + \Half (\bar J_+\bar J_+ + \bar J_- \bar
J_+) = J^2 $ in this representation.  ( A consequence of $J^2 = \bar
J^2$, for our leading order strong coupling Pomeron, is that we are
restricted to $n = h - \bar h = 0$ and are insensitive to rotations in
the impact parameter plane by $M_{12}$; we will learn more about this
below.)  So the boost Hamiltonian $H_{+-}$ is
\be
H_{+-}  = 3 - 2 J^2 - 2 \bar J^2
\ee
expressed in terms of $SL(2,C)$ Casimirs.  This equation realizes
the fact that the boost $M_{+-}$ commutes with all the generators
of $SL(2,C)$ in the conformal group.  
The quadratic form of the strong-coupling boost operator $M_{+-}$ 
then determines the $J$-plane eigenvalues to
quadratic order in $\nu$,
\be
j(\nu)= j_0 - {\cal D} \nu^2 + 0(\nu^4)  \; .
\ee
As first pointed out in Ref.~\cite{Brower:2006ea} the strong coupling
BFKL intercept is $j_0=2-2/\sqrt\lambda$ and the diffusion
constant\footnote{A cautionary note: in Ref.~\cite{Brower:2006ea} the
  integration variable used in solving the Pomeron equation
  (\ref{eq:jPomeron1}) is $2 \nu$ and this has the effect that the
  diffusion constant defined in Ref.~\cite{Brower:2006ea} is  ${\cal
    D}/4$ compared with the constant defined here.}  is ${\cal D} =
2/\sqrt\lambda$.

It is interesting to note that this structure is similar to the weak
coupling one-loop $n_g$ gluon BFKL spin chain operator in the large N
limit.  Here the boost operator is approximated by $M_{+-} = 1 -
(\alpha N/\pi) H_{\rm BFKL}$, where $H_{\rm BFKL}
= \frac{1}{4}\sum^{n_g}_{i=1}[{\cal H}(J^2_{i,i+1}) + {\cal H}(\bar
J^2_{i,i+1})]$, a sum over two-body operator with holomorphic and
anti-holomorphic functions of the Casimir. The Yang Mills coupling is
defined as $\alpha= g^2_{YM}/4\pi$. Even numbers of gluons ($n_g$)
contribute to the BFKL Pomeron with charge conjugations $C=+1$ and the
odd number of gluons to the so called ``odderon''~\cite{Lukaszuk:1973nt,Bialkowski:1974cp,
Kwiecinski:1980wb,Finkelstein:1989mf,Bartels:1999yt} with charge
conjugations $C=-1$.  To be more specific, the operator is defined by
its action on two body eigenstates~\cite{Kuraev:1977fs,Lipatov:1985uk,Braun:2003rp},
\be
{\cal H}(J^2) \Phi_{n,\nu} = \frac{1}{2}[\Psi(h ) + \Psi(1-h)- 2 \Psi(1)] 
\Phi_{n,\nu}.
\ee
The symmetry $h \rightarrow 1-h$ implies that this is a function of
$h(h-1)$ or equivalently the quadratic Casimir, which to first order is
\be \label{eq:BFKLleadingH}
{\cal H}(J^2)  + {\cal H}(\bar J^2)  \simeq  2\Psi(\half) - 2 \Psi(1) + 
\frac{1}{2}\Psi''(\half) [  J^2   +  \bar J^2  +1/2] \;.
\ee
Consequently in the two gluon channel with $n_g=2$ and $H_{\rm BFKL} =
\frac{1}{2}[{\cal H}(J^2_{1,2}) + {\cal H}(\bar J^2_{1,2})]$, the leading
eigenvalue with $n=0$ for the boost is given by
\be
j(\nu)= j_0 - {\cal D} \nu^2 + 0(\nu^4) \; ,
\ee
with $j_0 = 1 + 4\ln 2 \alpha N/ \pi $ and ${\cal D} = 14\zeta(3) \alpha N
/\pi $.  The two-gluon eigenvectors,
written in terms of complex transverse positions $b_i = x_i+iy_i$
for gluon $i$,
are 
\be
\Phi_{n,\nu}(b_1-b_0, b_2-b_0) = \big[ \frac{b_1 - b_2}{(b_1 - b_0)(b_2 
- b_0)}
\big ]^{i\nu + (1+n)/2} \;\big[ \frac{\bar b_1 - \bar b_2}{(\bar b_1 - 
\bar b_0)(\bar b_2 - \bar b_0)}
\big ]^{i\nu + (1-n)/2} \; .
 \ee
 They are given as a products of conformal and anti-conformal factors with
 weights $h = i \nu + (1+n)/2$ and $\bar h = i \nu + (1-n)/2$
 respectively.  Expanding in a Taylor series in $b_0$ and $\bar b_0$
 the wave function is easily seen as a sum of products of
 infinite-dimensional representations of a two-body Lie algebra.  This
 algebra has $\vec J_{1,2} = \vec J^{(1)} + \vec J^{(2)}$ represented
 by $J_0^{(i)} = b_i \dd_{b_i},J_-^{(i)} = - \dd_{b_i},J_+^{(i)} =
 b^2_i \dd_{b_i}$ and similarity for the antiholomorphic sector. In
 this representation the Casimirs are $J^2_{1,2} = -(b_1-b_2)^2
 \dd_{b_1} \dd_{b_2}$ and $\bar J^2_{1,2} = -(\bar b_1-\bar b_2)^2
 \dd_{\bar b_1} \dd_{\bar b_2}$.

 Let us note some differences between the strong-coupling and
 weakbrained-coupling limits.  First, $j_0$ moves from 1 to 2 as
 $\lambda$ moves from small to large.  Also, the formulas for $j(\nu)$
 given above have different regimes of validity; at strong coupling
 the energy-momentum tensor at $j=2$ (along with the nearby $j\sim 2$
 DGLAP operators) lies within the region of validity of the
 strong-coupling expression, while the explicit factor of $\lambda$ in
 $M_{+-}$ at weak coupling implies that Eq.~(\ref{eq:BFKLleadingH})
 breaks down before $j=2$.  Also, there is the fact that any $n$ is
 allowed at weak-coupling, but we see at strong coupling only $n=0$.
 Presumably this reflects the nearly point-like nature of a string in
 this limit, which leaves it unable to undergo rotation in the
 transverse plane within this approximation.  In strong coupling
 perhaps one should visualize the Pomeron as the exchange of single
 trace planar diagram with an infinite number of t-channel gluons
 whose interactions are treated via a mean field approximation.

\subsection{Pomeron Kernel at High Energies}
\label{subsec:obtainPomeron} 

With the $J$-plane Pomeron kernel, ${\cal K}(j, b^\perp, z,z')$,
expressed in terms of the $AdS_3$ propagator $G_3(j,v)$, we would also
like to examine the structure of a single Pomeron exchange at high
energies, ${\cal K}(s, b^\perp, z,z')$.  This is the kernel which
is used in the eikonal resummation, as reviewed in
Sec.~\ref{sec:eikonal}.

Given Eq.~(\ref{eq:ads3kernel}), it follows from 
Eq.~(\ref{eq:pomeronkernel1}) that the single Pomeron amplitude can be 
expressed,  after   wrapping the $J$-plane contour  to the 
left, as
\bea
 {\cal K}(s,b^\perp,z,z')   &=& -   (zz'/R^4)G_3(j_0,v)  \\
&\times& \widehat s^{j_0}
 \int_{-\infty}^{j_0}\frac {d j}{\pi }\; \frac{(1+ e^{-i\pi j}) }{\sin \pi j}\; \widehat s^{(j -j_0)}     \; \sin\left[ \xi(v)  \sqrt {2\sqrt{ \lambda}(j_0-j)}\;\right] \nonumber
  \label{eq:integraloverbfkl}
\eea
where we have exposed the dominant BFKL singularity at $j_0$. We have
also introduced $\xi(v)$ where $\cosh \xi = v+1$, in order to simplify
our expressions.\footnote{In terms of the new variable $\xi$, the
  combination $1+v + \sqrt {v(2+v)} = e^\xi$ and therefore $G_3(j,v)$
  also takes on a simpler form, $G_3(j,v)= e^{[2-\Delta_+(j)]\xi}
  /(4\pi\sinh\xi)$. }

There are two distinct high energy limits of interest to us: (1) $\log
\widehat s \rightarrow \infty$ with $\sqrt \lambda$ large but fixed
and (2) $\log \widehat s \rightarrow \infty$, $\lambda \rightarrow
\infty$ with $\log \widehat s /\sqrt \lambda \rightarrow 0$. The first
is the Regge limit which is dominated by the Pomeron exchange, and the
second is dominated by the graviton exchange.  Let us give an
approximate expression for ${\cal K}$ valid in both these regions.

We begin by separating ${\cal K}$ into its real and imaginary parts,
${\cal K} = {\rm Re}[{\cal K}] + i {\rm Im}[{\cal K}]$, 
\bea 
{\rm Re}[{\cal K}] &=& - (zz'/R^4)G_3(j_0,v) \widehat s^{j_0}
\int_{-\infty}^{j_0}\frac {d j}{\pi }\; \frac{(1+\cos \pi j) }{\sin \pi j}\; \widehat s^{j -j_0 }    
 \; \sin\left[ \xi(v)  \sqrt {2\sqrt{ \lambda}(j_0-j)}\;\right] \nonumber\\
{\rm Im}[{\cal K}] &=& (zz'/R^4)G_3(j_0,v) \widehat s^{j_0} \int_{-\infty}^{j_0}\frac {d
  j}{\pi }\; \widehat s^{j-j_0} \; \sin\left[ \xi(v) \sqrt {2\sqrt{
      \lambda}(j_0-j)}\;\right] \; .
      \label{eq:ImK}
\eea
With the change of  integration variable to  $y^2 = 2\sqrt{\lambda} (j_0 - j)$,
the imaginary part is recognized as a Gaussian integral that is
easily integrated exactly,
\bea
{\rm Im}[{\cal K}]  &=&   (zz'/R^4)G_3(j_0,v) \widehat s^{j_0}
 \int_{-\infty}^{\infty}\frac {d y}{2 \pi i \sqrt{\lambda} } \; y \; e^{ - \tau y^2/2\sqrt{\lambda}}     \; e^{i  \xi(v)  \; y }  \nonumber \\
&=&  (zz'/R^4)G_3(j_0,v)   ( \sqrt{ \lambda}/2\pi )^{1/2}    \xi \;  e^{j_0 \tau}\;  \frac{e^{- \sqrt{ \lambda} \xi^2/  2 \tau  }}{\tau^{3/2}} \; .
 \label{eq:imaginarypart}
\eea
The real part is more difficult. However we can find an approximation to
${\rm Re}[{\cal K}]$ that is uniformly valid for the region of interest,
where both $\log\widehat s$ and $\sqrt{\lambda}$ are large.
Large $\log\widehat s $ implies that the
cut in the $J$-plane is probed near the end point for $j-j_0 < O(1/\log\widehat s )$. Combined
with large  $\sqrt{\lambda}$, this allows us to expand the prefactor in $j-2$,
\be
\frac{(1+\cos \pi j) }{\sin \pi j} \simeq \frac{2}{\pi(j-2)} + O(j-2) \; .
\ee
The leading term implies the identity, $\dd_\tau [ e^{-2 \tau} {\rm Re}[{\cal K}] ] = - (2/\pi) e^{-2 \tau} {\rm Im}[{\cal K}] $ or  an approximation for the real part,
\be
{\rm Re}[{\cal K}]  \simeq  (zz'/R^4)G_3(j_0,v)    ( \sqrt{ \lambda}/2\pi )^{1/2}    \xi  \; \widehat s^2\; \int^\infty_\tau d\tau' \; \frac{2 e^{- 2 \tau'/\sqrt{\lambda} - \sqrt{ \lambda} \xi^2/  2 \tau'  } }{\pi \tau'^{3/2}} \; .
\label{eq:ReK}
\ee
Corrections are easily computed in a standard perturbation series. The
first order corrections to Eq.~(\ref{eq:ReK}) are  $O({\rm Im}[{\cal K}]/\log\widehat s )$ and
$O({\rm Im}[{\cal K}]/\sqrt{\lambda})$, but they are not needed in our present
analysis.

Let us first focus on the Regge limit: $\tau = \log \widehat s
\rightarrow \infty$ at fixed large $\sqrt{\lambda}$. In this limit the
end point dominates the integral in the expression (\ref{eq:ReK}) for
${\rm Re}[{\cal K}]$ and can be approximated by
\be\label{eq:approx1}
\int^\infty_\tau d\tau' \; \frac{2 e^{- 2 \tau'/\sqrt{\lambda} -
\sqrt{ \lambda} \xi^2/ 2 \tau' } }{ \pi \tau'^{3/2}} 
=
(\sqrt{\lambda}/\pi)\; \  e^{- 2 \tau/\sqrt{\lambda}}\ 
\frac{e^{- \sqrt{
\lambda} \xi^2/ 2 \tau } }{\tau^{3/2}} \left(1 + O(
\sqrt{\lambda}/\tau ) \right)
\ee
Combining this approximation for  ${\rm Re}[{\cal K}]$ with ${\rm Im}[{\cal K}]$, we have
the leading term in the Regge limit,
\be
{\cal K} \simeq (zz'/R^4)G_3(j_0,v) e^{j_0\tau} \left[(\sqrt{\lambda}/\pi)+ i\right] ( \sqrt{\lambda}/ 2 \pi )^{1/2}\; \xi  \; \frac{e^ {-\sqrt\lambda  \xi^2 / 2\tau}}{\tau^{3/2}}
\label{eq:transvasympbeh}
\ee
valid for $\sqrt \lambda /\log \widehat s \rightarrow 0$ and for
general value of $(\sqrt \lambda \xi^2)/\log \widehat s$.  This is our
key result.  Up to the log factors, the single Pomeron contribution in
a transverse representation at high energy is proportional to an
$AdS_3$ propagator at $j=j_0$ and a diffusion factor in $\xi$.   This
amplitude is complex, with its phase consistent with the Regge signature factor,
$(1+ e^{-i\pi j_0})/2 = (1+ e^{  2\pi i /\sqrt{\lambda}})/2 \simeq
1 + i \pi/\sqrt{\lambda}$, to leading order in $1/\sqrt{\lambda}$.

Before discussing in detail  the graviton limit, let us make a few
additional comments.  Let's return to momentum space, 
\be {\cal K}(s,t,z,z') = \int d^2b^\perp e^{-iq^\perp b^\perp} {\cal K} (s,b^\perp,z,z')\; , 
\ee
and examine the high energy behavior at fixed
$t$. With $t=-{q^\perp}^2\neq 0$, one easily verifies that our large
$\widehat s$ result, $ \widehat s^{j_0}/ \log^{3/2} \widehat s$ in
Eq.~(\ref{eq:transvasympbeh}), is consistent with a $\sqrt {j-j_0}$
BFKL singularity, as derived in \cite{Brower:2006ea}.  The limit
$t=0$, however, requires a more careful treatment. After a more
refined analysis, one can verify that Eq.~(\ref{eq:transvasympbeh})
leads to Eq.~(\ref{eq:ImKzero}) and is consistent with the
$1/\sqrt{j-j_0}$ singularity at $t=0$ found in
Ref.~\cite{Brower:2006ea}.

\subsection{Connection with Graviton Exchange}
\label{subsec:obtaingraviton}

Next we turn to the 
regime dominated by graviton exchange. 
For 
$\lambda \to \infty$, the dual theory approaches pure gravity, without
stringy corrections. For $\log \widehat  s
\rightarrow \infty$ and $\sqrt \lambda/\log \widehat  s \rightarrow \infty$,
the 
Pomeron exchange should smoothly become graviton exchange.  We recall that
the amplitude for the one-graviton-exchange Witten diagram in momentum
representation, for scalar sources on the boundary of $AdS_5$, is
\cite{Freedman}
\be
 {\kappa_5^2}\int  dz \sqrt g \int d z' \sqrt {g'} \; \tilde T^{MN}(p_1,p_3, z) \tilde G_{MNM'N'}(q,z,z') \tilde T^{M'N'}(p_2, p_4, z')  \label{eq:qonegraviton}
\ee
where $\kappa_5$ is the gravitational coupling in $AdS_5$, $\tilde
T^{MN}$ is the energy-momentum tensor for the scalar source in the
bulk and $\tilde G_{MNM'N'}$ is the graviton propagator, both in
momentum representation.  At high energies, keeping the leading
$++,--$ component, we have shown in Ref.~\cite{Brower:2007qh} that the
corresponding amputated amplitude in transverse representation is
\be
{\cal K}(s, x^\perp-x'^\perp,z,z') \sim { \widehat s}^2\; \left( \frac{zz'}{R^4}\right)   \;G_3(x^\perp-x'^\perp, z,z')  \label{eq:onegraviton}
\ee
where 
$G_3$ is the dimensionless scalar propagator for a particle of mass
$\sqrt{3}/R$ in an $AdS_3$ space of curvature radius $R$, and in the
conformal regime is equal
to the function $G_3(j=2,v)$ defined in Eq.~(\ref{eq:adschordal}).
We will now recover this from the Pomeron kernel.

First let us understand where the transition to this regime occurs.
For $\lambda$ sufficiently large,
the integral in the expression (\ref{eq:ReK}) for ${\rm Re}[{\cal K}]$ gets
its dominant contribution not from the end point at $\tau$ but at an
internal value, at the saddle point determined by $2
\tau'/\sqrt{\lambda} = \sqrt{\lambda} \xi^2/2\tau'$.  The
crossover between the Regge and the graviton  regimes is determined
by the collision of this saddle point with the end point,
\be
\xi = 2\tau/\sqrt{\lambda} = (2/\sqrt{\lambda}) \log \widehat s \; . \label{eq:GPtransition}
\ee

To determine the kernel in this regime, let
us return to the $J$-plane representation,
Eq.~(\ref{eq:integraloverbfkl}), obtained by closing the contour
around the BFKL branch point at $j_0$.
The
pole at $j=2$ in the integrand, although outside the integration
range, plays an increasingly important role in the limit $\lambda
\rightarrow \infty$. The dominant contribution to the
$J$-integral now comes from the region $j=j_0 - 0( 1/\sqrt\lambda)$
and the cutoff for the integral, instead of due to the $\widehat 
s^{(j-j_0)}$ factor, now comes from the last sine factor.  In terms of the variable $y=[2\sqrt\lambda(j_0-j)]^{1/2}$, the 
singularity at $j=2$ corresponds to poles at $y=\pm 2i$.

We first note that, due to the diffusion factor in Eq. (\ref{eq:imaginarypart}), ${\rm Im}[{\cal K}]$ vanishes in this limit. This is not surprising since this $j=2$ kinematic singularity does not contribute to ${\rm Im}[{\cal K}]$, (see Eq. (\ref{eq:ImK})),  and we therefore only need to focus on the real part, which can be expressed as Eq. (\ref{eq:ReK}). Changing integration variable to $\tau'=2 \tau/\sqrt \lambda$, the integral in Eq. (\ref{eq:ReK}) becomes
\be
\int^\infty_\tau d\tau\;  \frac{e^{- 2 \tau/\sqrt{\lambda} - \sqrt{ \lambda} \xi^2/  2 \tau  } }{\tau^{3/2}}=\sqrt 2\lambda^{-1/4}\int^\infty_{2\tau/\sqrt\lambda}  d\tau'\;  \frac{e^{-  \tau' -  \xi^2/   \tau'  } }{\tau'^{3/2}} 
\ee
In the limit $\tau/{\sqrt\lambda} \rightarrow 0$, the integral yields $\sqrt{2\pi} \lambda^{-1/4} \xi^{-1} e^{-2\xi}$, and
\be
{\cal K} \rightarrow {\rm Re}[{\cal K}] =\frac{2}{\pi}\widehat s^{2}\; \left(\frac{zz' }{R^4}\right)\; G_3(j_0,v)   e^{-2\xi (v)} =    \frac{2}{\pi}
\widehat s^{2} \;\left(\frac{zz' }{R^4}\right)\; G_3(2,v)  
\ .
\ee
This is the graviton result obtained in Ref.~\cite{Brower:2007qh}, where
$G_3(2,v)\equiv G_3(v)$, recovered through a $J$-plane analysis.

\section{Aspects of the Eikonal Representation}
\label{sec:unitarity}

In this section, we highlight several interesting features of the
eikonal approximation for multiple Pomeron exchange.  We focus
especially on issues of unitarity in the $s$-channel, and on the phase
of the amplitude, emphasizing its physical interpretation.  We also
discuss how particle-Pomeron amplitudes are embedded in our calculations, and
some connections with the eikonal approximation in flat-space string theory.

Of course the eikonal approximation is commonly used to obtain a
manifestly unitary amplitude for a study of unitarity. An S-matrix
element which can be approximated by an eikonal sum automatically
satisfies the unitarity bound as long as the imaginary part of $\chi$
is negative.  If the eikonal amplitude $\chi$ for elastic scattering
is real, then the eikonal sum precisely saturates the unitarity bound;
the elastic amplitude gives the total cross-section. Otherwise, the
imaginary part is related to inelastic processes not explicitly
described by the elastic amplitude.

Here, the situation is more subtle.  We want to compute
four-dimensional gauge amplitudes.  However, our methods involve the
{\it bulk} eikonal approximation, requiring us to integrate over bulk
coordinates, $z,z'$, for fixed $b$, as in Eq.~(\ref{eq:adseik}).  In
general, the bulk eikonal approximation is valid in only part of the
integration region, the ``eikonal region'' for short.  For this
reason, for most values of $b$, the gauge-theory amplitude cannot be
computed purely within the bulk eikonal approximation, and unitarity
of the gauge amplitude cannot be fully studied.  Nevertheless, as we
will see, unitarity of the bulk amplitude is still conceptually useful
and provides some important physical intuition. Later we will consider
large values of $b$ where the eikonal region makes the dominant
contribution to the gauge theory amplitude.

\begin{figure}[h]
\begin{center}
\includegraphics[width = 0.8\textwidth]{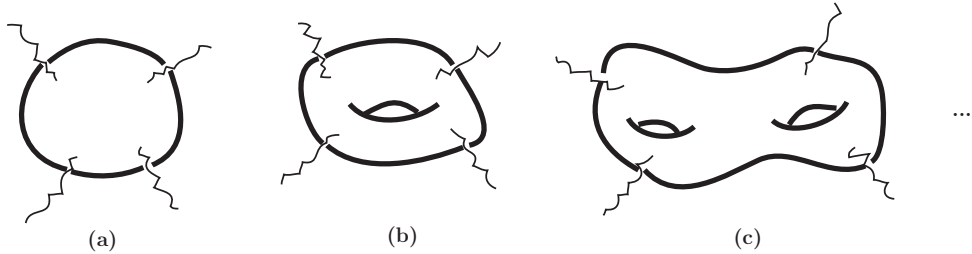}
\caption{Perturbative expansion for a four-point string amplitude: The
  planar approximation (a) has $s$-channel closed string excitation dual
  $t$-channel complex Regge exchange.  The torus diagram (b) has
  $s$-channel threshold for both single closed string excitations and a
  pairs of closed strings dual to Regge cuts. The two loop diagram gives
  one, two and three closed string production dual to Regge and multi-Regge
cuts, etc.}
\label{fig:loop}
\end{center}
\end{figure}

\subsection{Inelastic Production and $AdS$}

We begin by reviewing familiar aspects of
unitarity in four dimensions, and their extension in the present
context to physics in the relevant five bulk dimensions.

The unitarity condition
for the S matrix, $S^\dagger S=I$, can be
diagonalized through the $s$-channel partial-wave expansion.
The $2\to 2$ amplitude can be written
\be
A(s,t) = 16 \pi  \sum_l (2l+1) a_l(s) P_l(\cos \theta) \; ,
\label{eq:partialwave}
\ee
where $a_l(s)\equiv (s_l -1)/2i$, $s_l\equiv e^{2i\delta_l(s)}$.
The components of the diagonal
scattering matrix, $s_l= e^{2i\delta_l(s)}$, are unitary for a real phase-shift
$\delta_l(s)$.  In this case elastic scattering in this partial wave
saturates
unitarity.  More generally, if inelastic production is allowed,
$s^*_ls_l<1$, and the phase-shift $\delta_l(s)$ is complex,
with ${\rm Im}[ \delta_l(s)] > 0$.

For high-energy small-angle scattering, $t\simeq -(s/4) \theta^2$,
and with the identification $l \simeq b\sqrt s/2$, the sum becomes
approximately an integral over impact parameter $b$, so the
partial-wave expansion becomes approximately
\be
A(s,t) \simeq  -2i s \int d^2 b^\perp e^{-i q_\perp b^\perp}
\left( e^{2i\delta(s,b)} -1\right) \;
\label{eq:partialwavehighs}
\ee
For a real phase shift $\delta(s,b^\perp)$ at high energy, unitarity for the
the transverse amplitude,
\be
\widetilde A(s,b^\perp) =\int \frac{d^2q_\perp}{ (2\pi)^2 } e^{ ib^\perp
q_\perp}  A(s,t),
\ee
becomes a scalar condition:
\be
 {\rm Im}\;  \widetilde A(s,b^\perp) =(1/4s)  | \widetilde A(s,b^\perp)
|^2\; .
 \label{eq:4dlocalunitarity}
\ee

In general, at large $s$, the on-shell amplitude 
is an integral of an off-shell position-space Green's function, which
depends on the four transverse positions of the two incoming and two
outgoing particles.  Three of these transverse-position variables are
independent; the fourth is removed by translation invariance.  In the
eikonal approximation, the scattering amplitude at each order in $\chi$ is
proportional to a product of $\chi$ and two delta functions, $\delta^2(x^\perp_1-x^\perp_3)
\delta^2(x^\perp_2-x^\perp_4) $, 
which ensures that neither scattered particle is deflected by
a transverse shift in position.  This effect of
``zero-transverse-deflection'' removes two more of the
transverse-position variables, leaving an eikonal kernel $\chi$ that
is a function of only one.  When the eikonal amplitude is
exponentiated, the effect of zero-transverse-deflection is to ensure
the full amplitude remains a function of only one transverse position
variable.  If the eikonal approximation is valid for a given partial
wave, that is for a given $b^\perp$ (at fixed $s$), then the eikonal kernel
is nothing but the phase-shift, $\chi(s,b^\perp) =
2\delta(s,b^\perp)$.  The general requirements of unitarity on the
phase shift $\delta$, partial wave by partial wave, thus descend to
requirements of unitarity on $\chi$ which are local in $b$.  For this
reason, we can cease to worry as to whether the eikonal approximation
applies to the entire $S$ matrix; it is enough for us that there are
some partial waves to which it applies.

Up to this point we have been discussing standard ideas in four
dimensions.  Now we turn to the calculation which we have addressed in
this paper, which has so far been performed only in conformal
four-dimensional field theories (a condition which we will relax later, 
but which we may retain for now.)  Although the
conformal theory has no S-matrix, this is not relevant, since we can
add a heavy quark as a probe of the conformal theory and study
onium-onium scattering.  More important, most gauge theory amplitudes
cannot be computed fully within the eikonal approximation; only in
some regions of the scattering phase space can it be used.  But the
eikonal approximation allows us to apply notions of unitarity locally
in five dimensions.

To make this statement precise would require a generalization of the
notion of partial waves, as in Eq.~(\ref{eq:partialwave}), to the
bulk.  We sidestep this by noting that our form of the amplitude is
already a generalization of its high-energy limit
Eq.~(\ref{eq:partialwavehighs}). 
Thus we 
should again view $\chi(s,b^\perp, z, z')$ as
proportional to the phase shift in the high-energy 
limit of the bulk partial wave expansion,
\be
\; A_{2\to2}(s,t) \simeq
\int d^2b \; e^{-ib^\perp q_\perp} \int dz dz' P_{13}(z) P_{24}(z')
\widetilde A(s,b^\perp,z,z')\;,
\ee
\be
\widetilde A(s,b^\perp,z,z')= -2 i s \left [ e^{i\chi(s,b^\perp, z,z')} -
1\right] \;.
\ee
The key difference between this and the previous case is that the $z$
coordinate is not translationally invariant, with two important
consequences.  First, the bulk
amplitude in transverse position space is a function of four $z$
coordinates, which in the eikonal approximation are reduced to two ($z$
and $z'$), as in Eq.~(\ref{eq:eikonalcontraction}).  Second, the wave functions for the incoming and outgoing
particles are not simply plane waves in the $z$ coordinate, and
instead of the simple factor $e^{-ib^\perp q_\perp}$ which is left
over from the wave functions in the Minkowski coordinates, we have the
more complicated products of wave functions $P_{13}(z) P_{24}(z')$,
defined in Eq.~(\ref{eq:overlapfunction1}).

Just as the eikonal approximation may not apply in an ordinary
four-dimensional scattering amplitude, but may apply in certain
partial waves, so here the eikonal approximation will apply only in
the limited region we called the ``eikonal region''.  Where it does, the
full amplitude can be expressed through $\widetilde
A(s,b^\perp,z,z')$, a function of one relative Minkowski coordinate
and two bulk radial positions, and unitarity applies to it as before.
\be
 {\rm Im}\;  \widetilde A(s,b^\perp,z,z') \geq (1/4s)  | \widetilde
A(s,b^\perp,z,z') |^2  \; .
 \label{eq:localunitarity}
\ee
 A real amplitude $\chi$, as for the pure gravity case, saturates the
bound, while
the complex amplitude of the Pomeron $\chi$ satisfies a corresponding
inequality.

\subsection{Physical Consequences}
\label{sec:physical}
We now consider the meaning both of the imaginary part of $\chi$ and of
the imaginary part of $-2is(e^{i\chi}-1)$.  Let us expand the eikonal sum in
powers of $\chi$:
\bea\label{expandinchi}
{\rm Im} \widetilde A(s,b,z,z')&=& - 2s\; {\rm
Re}\left[\sum_{n=1}(i\chi(s,b,z,z'))^n /n!\right]
\nonumber \\
&=&  s  \left\{ 2\; {\rm Im} [ \chi(s,b,z,z')] +  {\rm Re} [ \chi^2(s,b,z,z')]
+ \cdots\right\}
\;.
\eea
If $\chi$ is real, as in graviton exchange, contribution to ${\rm Im} \widetilde A$ begins 
at one-loop; if $\chi$ is complex, as for the Pomeron, there is a
tree-level contribution.

Whereas the magnitude of the eikonal, (up to $\log $ factors),  grows as
$G_3(b^\perp, z,z') {\widehat s}^{j_0-1}$,
its phase is a {\it constant}, depending only on $j_0$.  From
Eqs.~(\ref{AdSpomeronchi}) and (\ref{eq:integraloverbfkl}),
 \be\label{chiphase}
 \chi(s,b,z,z')   \simeq  e^{i(1-j_0/2)\pi} |\chi(s,b^\perp, z,z') |\;.
  \ee
Recall $j_0$ is the intercept  of the leading $J$-plane
singularity, which moves from $j_0 \simeq 1$ at weak coupling to $j_0
\simeq 2$ at strong coupling. This expression requires large energy, so 
it fails at small $z,z'$ where the locally measured 
center-of-mass energy is comparable
to or less than the string scale; there the phase goes to zero, for reasons to
become clear in a moment.

From the perspective of a $J$-plane
analysis, an eikonal sum represents an approximate treatment for
multi-Pomeron $J$-plane singularities, with the $n$-loop diagram giving
rise to an $n$-Pomeron cut at $j = n (j_0 -1) + 1$. For example, the
one-loop diagram grows like $s^{2j_0-1}$ and
has a two-Pomeron cut at $2j_0-1$.  Were $j_0 < 1$,
higher order contributions would be subleading and the eikonal sum would be a
rapidly convergent expansion at large $s$, but this is not the case for $1 <j_0
< 2$, the range of interest here. Therefore, where the eikonal expansion
is reliable, it is interesting to discuss the relative importance of the
various
perturbative contributions
to the absorptive part of the forward amplitude, ${\rm Im} [A(s,0)]$,
and, thus to the total cross section $\sigma_{total}(s)$, through the
optical theorem, $\sigma_{total}(s) \simeq (1/s) {\rm Im} [A(s,0)]$.

It is important first to elucidate the physical meaning of the phase.
In potential theory, elastic scattering dominates when $\chi$ is real.
On the other hand, a black disk (total absorption) gives a
pure-imaginary $\chi$.  A potential with a real and an imaginary part
gives a complex $\chi$, corresponding to partial
absorption.\footnote{For a short-range potential with non-vanishing
real and imaginary parts,
$
V(s, r)=  V_R(s,r)  - i V_I(s,r)
$
the eikonal at high energy is given by
\be
      \chi (s, b^\perp) = - (2\mu/\sqrt s) \int_{-\infty}^\infty dz \;
V\left(s, \sqrt{{b^\perp}^2+z^2}\right)
\ee
where $r^2=x^2+y^2+z^2={b^\perp}^2+z^2$ and $\mu$ is the particle
mass.  (See, for instance, Schiff, Quantum Mechanics, page: 339-343.)}
At weak 't Hooft coupling, as in QCD itself, Pomeron exchange
corresponds to $j_0\to 1$, for which ${\rm Re}[\chi] \to 0$; there is
nearly complete absorption, and Pomeron exchange approximates a black
disk.  For graviton exchange, $j_0=2$ and there is no absorption, but
at finite large 't Hooft coupling $j_0= 2- 2/\sqrt\lambda$ the
imaginary part of the tree-level amplitude is nonzero, ${\rm Im}\;[2
\chi] \propto {1\over \sqrt\lambda}$.  
This absorptive
part arises from averaging the effect of massive $s$-channel string
resonances that arise in the tree-level string amplitude,
Fig.~\ref{fig:loop}a.  A cut across an exchanged Pomeron looks like a
massive string in the $s$ channel, which gives a pole if $s$ is equal
to the mass of a string; averaging over these poles at large $s$ gives
the Pomeron exchange a nontrivial phase.  This inelastic process
represents a form of absorption out of the two-string Hilbert space
and into the one-string Hilbert space.  Other forms of absorption
cannot contribute at leading order in the string coupling and at the
leading power of $s$.\footnote{This type of cut across a large closed
string representing, in a confining field theory, a highly excited state that
will decay to many hadrons, is called an AGK cut~\cite{Abramovskii:1972zz,Abramovsky:1973fm}. }

When we move to the one-loop graph, a number of interesting issues
arise.  On the one hand the imaginary part of the one-loop amplitude
is proportional, in the expansion of $2is [e^{i\chi}-1]$, to ${\rm Re}
[\chi^2]$.  On the other hand, this same quantity should be given by
looking at all the cuts through the one-loop graph, of which there are
several.  If $\chi$ is real, as in the exchange of an elementary
particle such as a graviton, then the only cut is the obvious one,
cutting through both of the scattered particles.  This cut is
proportional to $\chi^*\chi$, which is obviously equal to ${\rm Re}
[\chi^2]$ in this case.

If $\chi$ is complex, then one might at first glance add to this first
cut two more cuts, one through each Pomeron, which summed together
give
\be
 i \chi {\rm Im}\; [2  \chi] +  {\rm Im}\; [2 \chi] (i\chi)^\dagger
= - ({\rm Im}[2\chi] )^2 \ .
\ee
 That this is insufficient can be seen by
considering the following identity,
\bea\label{chichiidentity}
&& {\rm Re}\; [  \chi(s,b,z,z')  \chi(s,b,z,z')] \nonumber\\
&=&  \chi^\dagger(s,b,z,z')  \chi(s,b,z,z') + i \chi(s,b,z,z') {\rm Im}\;
[2  \chi(s,b,z,z')] \nonumber \\
& & +  {\rm Im}\; [2  \chi(s,b,z,z')] (i \chi(s,b,z,z'))^\dagger +
\frac{1}{2!} ({\rm Im}\;[2  \chi(s,b,z,z')])^2
\label{eq:4cuts}
\eea
which, from Eq.~(\ref{chiphase}), reads
 \be\label{chichiphases}
 \cos (j_0 \pi) |\chi |^2 = \left[ 1 - 2  \sin2(j_0\pi/2) -
2  \sin2(j_0\pi/2) +2  \sin2(j_0\pi/2)\right] |\chi |^2
\ee
The left hand side is the contribution to ${\rm Im}[ \tilde A]$ from the second term in $- ie^{i\chi}$;
the first (positive) term on the right
is the cut through the scattered particles, and the second and third
(negative) terms are the cuts through the left and right Pomeron
respectively in Fig.~\ref{fig:loop}b.  Consistency with Eq.~(\ref{expandinchi}) requires the
last term must be present.  It arises in string theory from cutting
the torus diagram as one would slice a bagel, with the incoming states
on one slice and the outgoing on the other.\footnote{Long ago, in
Regge theory, this contribution was identified within field theory.
Corresponding to a ``Mandelstam diagram''~\cite{Mandelstam:1963a,Mandelstam:1963cw}, it is the
essential mechanism to generate $j$-plane cuts from the exchange of
two Regge poles.}  This positive contribution to the imaginary part
corresponds to a new on-shell process not yet included: 2 massive
strings propagating in the $s$-channel.  Only with all four cuts do we
obtain the correct second term in Eq.~(\ref{expandinchi}).\footnote{In
string theory there are really only three cuts when viewed topologically, but the first and
fourth term above become independent cuts in the limit in which we are
working, where we separate massless closed string states in the
$s$-channel from massive ones.  
Corresponding subtleties also arise at
higher loops. Phenomenologically, this separation identifies ``diffractive" vs ``non-diffractive" production. 
Here, diffractive production refers to final states having a large rapidity gap. }  

This feature generalizes: for the $n$-loop amplitude, one finds
$2^{n+1}$ terms corresponding to up to $n+1$ strings propagating in
the $s$ channel.  In fact all the statements made here for the
two-Pomeron exchange graph generalize to all orders, through
straightforward combinatorics that build up the exponential.

It is interesting to compare these features with those arising in the
QCD literature regarding the phases in single and multiple Pomeron
exchange processes.  We have just seen that five-dimensional Pomeron
exchange, at leading order, gives a four-dimensional amplitude which
is proportional to ${\rm Im}[s^{j_0}]$, where $j_0$ is slightly less
than 2.  QCD data has long been modeled~\cite{Capella:1992yb} with a
similar single four-dimensional Pomeron exchange with $j_0$ slightly
larger than 1.  In both cases this poses a problem, since the total
cross section, proportional to the imaginary part of the forward
amplitude, grows too fast to be consistent with unitarity.  In QCD one
remedy has been to consider the correction from two-Pomeron exchange,
which goes as $s^{2j_0-1}$, and whose imaginary part is {\it
negative}, from the left-hand side of Eq.~(\ref{chichiphases}).  This
correction therefore gives a negative contribution to the growing
single-Pomeron total cross-section.  But for $j_0>1.5$, this picture
cannot survive, since the imaginary part of the two-Pomeron exchange
correction is positive.  Thus, while it is often sufficient at weak
coupling to treat the absorptive correction to Pomeron exchange by
keeping only the two-Pomeron cut, in strong coupling the totality of
the whole eikonal sum must become important.\footnote{That the
character of diffractive scattering should change as one moves from
the region $j_0<1.5$ to $j_0>1.5$ was noted in
Ref.~\cite{Brower:2006ea} in comments relating to black hole
production.}

\subsection{A Multi-Channel Interpretation}
\label{sec:multichannel}

We have, up to now, discussed $\chi(s,b,z,z')$, and considered phases
and unitarity as applied locally in the bulk.  But it is interesting
to return to the four-dimensional gauge theory, and to consider what
our current discussion means in that context.  In particular, in those
limited computations where the eikonal region includes the entire
bulk, it is possible to reinterpret the bulk eikonal amplitude, a
function of $z$ and $z'$, as a field theoretic eikonal amplitude which
is a matrix representing transitions between Kaluza-Klein modes.  That
is, if $n_1,n_2$ are Kaluza-Klein modes which scatter into modes
$n_3,n_4$, the amplitude for that transition is a matrix, which is
itself the exponential of a matrix eikonal kernel $\widehat\chi$:
\be
- 2is\left\{\exp\left[i\widehat\chi\right] - 1 \right\}_{n_4,n_3;n_2,n_1}
\ee
This represents a multi-channel eikonal approximation, which one
could have attempted from the start within quantum field theory.  From
such a point of view, it would hardly be obvious that the matrix
$\widehat\chi$ could be simply diagonalized by representing the modes labeled
by $n_i$ as functions on a new bulk $z$ coordinate.  In this sense, the
gauge-gravity duality performs a small miracle.

Technically, this issue is most easily discussed in the presence of a
discrete hadron spectrum, but this requires more formalism than we
have presented here.  Instead we will simply regulate our conformal field
theory with a hard infrared cutoff, which is for many purposes
effectively the same thing.  We
temporarily introduce an IR cut-off in the $AdS_5$ space,
$0<z<z_{IR}$. The $AdS_5$ spectrum is now discrete, consisting of an
infinite sequence of stable KK modes with normalized orthogonal wave
functions, $\Phi^{(n)}(z)$,
\be
\int_0^{z_{IR}} dz \sqrt g (z/R)^{2} \Phi^{(n)}(z) \Phi^{(m)}(z) = \delta_{n,m}
\label{eq:orthonormal}
\ee
Instead of enumerating them by the co-ordinate $z \in [0,z_{IR}]$, we
can change basis to the physical on-shell scattering states, using the
completeness relation,
$
\sum_n \Phi^{(n)}(z) \Phi^{(n)}(z') = (z/R)^3 \delta(z-z').
$
In this basis we have a matrix eikonal expression for all the 2-to-2
on-shell scattering amplitudes,
\be
A_{n_4,n_3 \leftarrow n_2,n_1}(s,t) = - 2 i s \int d^2b e^{-i b q_\perp} \;
\left [ e^{i\widehat \chi(s,b)} - I \right]_{n_4,n_3;n_2,n_1}  \; ,
\ee
where $\widehat \chi$ is a matrix for all possible 2-to-2 scattering
amplitudes with a single Regge exchange kernel,
\be
\chi_{n_4n_3;n_2 n_1}(s,b) = \int\ dz\ dz'\ P_{n_3n_1}(z) P_{n_4n_2}(z')
\chi(s,b, z,z')
\ee
with $ \chi(s,b, z,z')$ given by Eq.~(\ref{AdSpomeronchi}), and
$P_{ij}$ are ``overlap functions'', defined in
Eq.~(\ref{eq:overlapfunction1}).   Note that the
eikonal matrix is symmetric,
$\chi_{n_4n_3;n_2n_1}(s,b)=\chi_{n_2n_1;n_4n_3}(s,b)$. If $\chi(s,b,z,z')$ is real, the
eikonal matrix is hermitian,
$\chi_{n_4n_3;n_2n_1}(s,b)=\chi^*_{n_2n_1;n_4n_3}(s,b)$, hence the
theory is unitary after incorporating all 2-body inelastic channels
made of KK modes.  The $s$-channel unitarity now takes on a matrix
form,
\be
{\rm Im}\; A_{n_4n_3; n_2n_1}(s,b^\perp) =(1/4s)  \sum_{n,m} 
A^\dagger(s,b^\perp)_{n_4n_3;nm}   A(s,b ^\perp)_{n m; n_2n_1}
\label{eq:multichannelelastic}
\ee

From the field theory point of view, it is remarkable that the scattering
matrix of the KK
modes can be diagonalized by introducing a single geometric
co-ordinate $z$.
The
eikonal amplitude in this basis leads directly back to
Eq.~(\ref{eq:adseik}).  Indeed, using the orthonormal condition,
Eq.~(\ref{eq:orthonormal}) and the associated completeness relation,
one can convert the multi-channel unitarity condition,
Eq.~(\ref{eq:multichannelelastic}), into the local unitarity
condition, Eq.~(\ref{eq:localunitarity}), with equality if $\chi$ is real and inequality if ${\rm Im}\chi >0$.

Finally we may let $z_{IR}$ go to infinity; the modes become continuous
but the above relations survive unchanged.
Local elastic unitarity
remains meaningful when the IR cut-off is removed. In the conformal
limit this diagonalization may be viewed a consequence of a conformal partial
wave expansion as described in Ref.~\cite{Braun:2003rp}. However, it is
more general, and applies in nonconformal theories.

The multichannel expression in Eq.~(\ref{eq:multichannelelastic})
includes the effects of the Kaluza-Klein modes of the bulk graviton,
but not those of the higher-mass (and higher-spin) string states.  One
might hope that this expression can be generalized to include them,
and even that the simplicity of the bulk eikonal phase might
generalize to the full string theory.  While we have not shown this,
we will note in Sec.~\ref{sec:ACV} that there are interesting and suggestive
simplifications in the eikonal approximation to scattering of
flat-space strings.  It remains to be seen how to incorporate these
into a fuller understanding of eikonal scattering of strings in curved
space, and whether the difficulties of quantizing strings in
Ramond-Ramond backgrounds can be evaded.

\subsection{Two-Pomeron Cut and Particle-Pomeron Amplitude}
\label{sec:twopomcut}

The strength of the two-Pomeron cut, which we have evaluated in
Sec.~\ref{sec:eikonal}, can be interpreted as proportional to the
square of the ``fixed-pole'' residue of the particle-Pomeron
amplitude. This then provides a connection to the discussion of closed
string eikonalization of Ref.~\cite{Amati:1987uf}, which we will turn
to in Sec.~\ref{sec:ACV}, and also allows a generalization of our
treatment to include additional corrections such as triple-Pomeron
interactions which have been left out of our current analysis.

We have pointed out in Sec.~\ref{sec:physical} that an eikonal sum
provides an approximate treatment for the strengths of multi-Pomeron
$J$-plane singularities. In a $d=4$ field theoretic setting, the
existence of multi-Pomeron $J$-plane singularities was demonstrated by
a careful analysis of the analytic structure of non-planar Feynman
graphs~\cite{Mandelstam:1963a,Mandelstam:1963cw}. This can be generalized to strings through
the notion of fixed-pole residues for particle-Pomeron amplitudes. As
explained in Sec.~\ref{sec:fixedpole}, a fixed-pole residue at $J=-1$
corresponds to extracting certain  spectral weight for an analytic
function, e.g., Eq.  (\ref{eq:fixedpoleresidue}). For this residue not
to vanish, the amplitude must satisfy an unsubstracted dispersion
relation and has both left- and right-hand cuts.

  Recall that 
in our one-loop analysis in Sec.~\ref{sec:eikonal}, we
are left with two integrals, $\int d q_1^- A_{13}$ and $\int d q_1^+
A_{24}$, where $A_{13}$ and $A_{24}$ are particle-Pomeron amplitudes.
Let us focus on the integral over $A_{13}$, which, in strong
coupling, is the a sum of two $AdS_5$ scalar propagators, $G_5$,
Fig.~\ref{fig:pompart}. As a function of $M^2=2 p_1^+q_1^-$, $A_{13}$
has both a right-hand cut, from the $s$-channel propagator, and a
left-hand cut, from the $u$-channel propagator. It can be represented
through a dispersion relation, Eq.~(\ref{eq:DR}), as the sum of two
analytic functions, $A_{13}(M^2)= A_R(M^2)+ A_L(M^2)$, with a right- and
left-hand cut in $M^2$ respectively.  Up to a common factor of
$R^{-3}$, the first term, $A_R$, is $G_5(M^2)$ and the second, $A_L$,
is $G_5(-M^2)$, where we have suppressed their dependence on
transverse coordinates.  It can easily be shown that $G_5(M^2) =
0(1/M^2)$, whereas the sum $A_{13}(M^2)$ is even in $M^2$, from which
we deduce that $A_{13}(M^2)=0((1/M^2)^2)$ for $|M^2|\rightarrow \infty$.

From the one-loop integral, the path of $q_1^-$-integration goes over the
$s$- and $u$-channel physical regions of the amplitude $A_{13}$. It leads to an integral in $M^2
$ which goes over the right-hand cut and under the left-hand cut of
$A_{13}(M^2)$. Since $A_{13}$ vanishes at infinity as $0((1/M^2)^2)$,
the integration contour can be freely rotated into the complex $M^2$ plane, and the integral $\int d q_1^-  A_{13}$ , 
after multiplying by $p_1^+/(i\pi)$, becomes 
\be
\frac{1}{2\pi   i }  \int_{-i\infty}^{i\infty} dM^2 A_{13}(M^2) 
\ee
Note that the integral is precisely of the form
Eq.~(\ref{eq:fixedpole}), discussed in Sec.~\ref{sec:fixedpole}, and
it is the $j=-1$ fixed-pole residue for the particle-Pomeron
amplitude, $A_{13}$. Here, $G_5(M^2)$ and $G_5(-M^2)$ play the role of
one-sided analytic functions $A_R$ and $A_L$ respectively.

 The spectral representation for $G_5$,
and a completeness relation for Bessel functions, leads to 
\be
\frac{1}{2\pi   i }  \int_{-i\infty}^{i\infty} dM^2 A_{13}(M^2) =  C_{13} =    \delta^2 (x_1^\perp- {x_3}^\perp) (z/R)^3 \delta(z_1-z_3)
\ee
where we have put back the dependence on transverse coordinates. Together with a similar 
integral over $ A_{24}$, it leads to a remarkable result
\be
\frac{ 1}{\pi ^2}\int dq_1^- A_{13} \int dq_1^+ A_{24}=  (2/s)   C_{13}   C_{24}=  (2/s)(zz'/R^2)^{-2}\< 3,4|1,2\>
\ee
from which Eq.~(\ref{eq:eikonalcontraction2}) follows.
Since this corresponds to the contribution for the 2-Pomeron cut, we
see that the strength of this cut is proportional to the product of
two fixed-pole residues for the respective particle-Pomeron amplitude.

\subsection{Frozen String Bits in Flat Space}
\label{sec:ACV}

It is interesting to compare our strong coupling results in $AdS$ space
with the eikonal formula of Amati, Ciafaloni and
Veneziano~\cite{Amati:1987wq,Amati:1987uf} for the superstring in flat
space. The flat space solution does not require a truncation of the
infinite number of normal modes of a full string world sheet
description, so similarities with the general mechanism for
eikonalization in string theory found in our strong coupling AdS
example suggest further generalization beyond strong coupling. In flat
space the superstring eikonal phase $\widehat \chi$ is a matrix for
all 2 to 2 particle scattering amplitudes in the planar approximation.
Similarly to our multichannel eikonal amplitude, this matrix can be
re-expressed geometrically, this time by a change of basis to an
infinite dimensional ``impact parameter'' space for the transverse
positions of individual string ``bits'' $x_\perp(\sigma)$ of the
colliding strings.

Let us review a few of the results of Refs.~\cite{Amati:1987wq,Amati:1987uf}.
Consider
the eikonal approximation for graviton-graviton
elastic scattering.  The first term is the
Regge approximation to the planar diagram for graviton scattering,
\be
A(s,t) =   (\epsilon_3 \cdot \epsilon_1) (\epsilon_4 \cdot \epsilon_2) g^2_s {\cal K}_{\cal P}(s,q_\perp) \; ,
\label{eq:planar}
\ee
where $\epsilon_i$ are the graviton polarization tensors. The kernel
for a $t$-channel Reggeized graviton exchange is 
\be
 {\cal K}_{\cal P}(s,q_\perp) = 2   \frac{\Gamma(1- \alpha(t)/2)}{\Gamma(\alpha(t)/2)} (  e^{-i \pi/2} \alpha' s/4)^{\alpha(t)} \; ,
\label{eq:Pkernel}
\ee
where $t = - q^2_\perp$,  $\alpha(t) =2 + \alpha' t/2$.

As before, the key step for the eikonal
approximation of each term in the expansion is well illustrated by the
one-loop diagram. In the high-energy small-angle limit, this diagram
can be expressed as a box diagram,
Fig.~\ref{fig:pompart}, with two Pomeron exchange kernels
(\ref{eq:Pkernel}) for the rungs, coupled to the 2-to-2
Pomeron-graviton scattering amplitudes, $A_{13}$ and $A_{24}$, on
the sides of the ladder.  Indeed for the flat space superstring, the
legitimacy of this approximation has been demonstrated by 
Sundborg~\cite{Sundborg:1988tb} through a detailed
analysis of the high energy limit of the exact one loop superstring
diagram.

As a result it is proven~\cite{Amati:1987wq,Amati:1987uf} 
that the leading two Pomeron cut contribution reduces
to the analysis of the Pomeron-particle scattering amplitudes
\be
A_{13}(M^2,q_1,q_2) = \int \frac{d^2z}{\pi} |w|^{- 2 - \alpha'M^2/2} |1-w|^{\alpha' q_1 q_2} 
\ee
on the left side of the ladder, 
and similarly for $A_{24}(M'^2,q_1,q_2)$ on the right
side, as illustrated in Fig.~\ref{fig:box}. Here we
have defined $M^2 =- (p_1 + q_1)^2 $, $M'^2 =- (p_2 - q_1)^2$ and
$\alpha' q_1 q_2 = \alpha'( t_1 + t_2 - t)/2$. The box diagram is
evaluated by rotating the contour in $M^2$ for $A_{13}$ (and in $M'^2$
for $A_{24}$) in direct analogy with our discussion in
Sec.~\ref{sec:twopomcut}. Parameterizing the world sheet by $w =
\exp[\tau + i\sigma]$, the result is
\be
C_{13} = \frac{1}{4 \pi i} \int^{i \infty}_{-i \infty}  dM^2 A_{13}(M^2,q_1,q_2) =
\int^{2 \pi}_0 \frac{d\sigma}{2 \pi} |1 - \exp(i \sigma)|^{\alpha' q_1 q_2}
\label{eq:acv_fp}
\ee
and similarly $C_{24}$. As noted above, these integrals define the
residue of the $J= -1$ fixed pole for a Pomeron-particle scattering
amplitude, which sets the strength of the two Pomeron cut at $2 j_0
-1$. 

Now to make the comparison with our $AdS$ eikonal result, it is useful
to recast the derivation of Ref.~\cite{Amati:1987uf} in light-cone
gauge~\cite{Giddings:2007bw,Brower:2006ea}.  In light-cone gauge, by
suitable worldsheet diffeomorphism, the longitudinal modes are fixed
so that $X^+(\sigma,\tau) = \tau$ and $X^-(\sigma,\tau)$ is dependent
on transverse modes via the Virasoro constraint.  There are similar
conditions on the world sheet fermions, although they don't contribute
to the leading eikonal limit. The result is that the transverse
coordinates of the string in target space, $X_\perp(\tau,\sigma)$,
form a complete set of independent bosonic degrees of freedom. One
consequence is that the fixed pole integral, Eq.~(\ref{eq:acv_fp}),
demonstrates again that the leading contribution to the eikonal
approximation of the one-loop diagram is instantaneous in target space
light-cone time: $x^+ = \tau$.  Moreover, as demonstrated in
Ref.~\cite{Amati:1987uf}, this observation holds order by order.  For
the $n$-Pomeron exchange graph, the leading contribution to the
$n$-Pomeron cut in the $J$-plane, which is of order $~(g^{2n}_s
s^{n(j_0-1)+1})$, is instantaneous in $x^+$.

Next, following steps similar to our eikonal derivation above,
one arrives at the eikonal amplitude~\cite{Amati:1987wq,Amati:1987uf},
\be
T_4(s,t)=  -2i s (\epsilon_3 \cdot \epsilon_1) (\epsilon_4 \cdot \epsilon_2) \; \int d^{D-2}b_\perp e^{i b_\perp q^\perp} \;\langle 0;0| [ e^{i \widehat \chi(s,b_\perp;\hat X_\perp, \hat X'_\perp)} -1] | 0; 0  \rangle 
\ee
with the matrix  phase,
\be
 \widehat \chi(s,b_\perp;\hat X_\perp, \hat X'_\perp) = \frac{g^2_s}{2s} \int \frac{d^{D-2}q_\perp}{(2 \pi)^{D-2}} {\cal K}_{\cal P}(s,q_\perp) \int \frac{d\sigma d\sigma'}{(2\pi)^2}
e^{\textstyle i q_\perp [b_\perp + \hat X_\perp(\sigma) - \hat X'_\perp(\sigma')] } \; .
\ee
The state $|0;0 \rangle$ is the string vacuum~\cite{Polchinski:1998rq} and
$\hat X_\perp(\sigma)$ are the non-zero mode transverse position operators for 
the string.\footnote{At $|w| = 1$  or $\tau = 0$, the zero mode
$\hat x^\perp_0 = \int d\sigma \hat X_\perp/2 \pi$ gives the
delta function $\delta^{D-2}(p^\perp_1+p^\perp_2+p^\perp_3+p^\perp_4 )$.}

Here we also note that one can exactly diagonalize this matrix by changing
basis from the eigenstate of the light-cone string Hamiltonian,
\be
P^- = \frac{1}{2p^+} \oint  d\sigma [(\hat \Pi_\perp(\sigma))^2 + \frac{1}{(2 \pi \alpha')^2} 
(\dd_\sigma \hat X_\perp(\sigma))^2]
\ee
to the string bit basis, $|x_\perp(\sigma) \>$, that diagonalizes the
transverse position operators $\hat X_\perp(\sigma)$. In light-cone
gauge, both bases are a complete representation of all the physical
bosonic (non-spurious) modes of the superstring.  We change basis for
both the right-moving string $x_\perp(\sigma)$ and the left-moving
string $x'_\perp(\sigma')$, obtaining
\be
T_4 \sim -2i s \; \int  {\cal D}x_\perp {\cal D}x'_\perp d^{D-2}b_\perp  
P_{13}[x_\perp(\sigma)]  P_{24}[x'_\perp(\sigma')]  e^{i b_\perp q^\perp} 
\big[ e^{i \chi(s,b_\perp; x_\perp,  x'_\perp)} -1\big] \; .
\ee
The string bit probability distributions for flat space string theory
\be
P_{31}[x_\perp(\sigma)] = |\Phi[x_\perp(\sigma)]|^2  \tbox{and}
P_{42}[x'_\perp(\sigma')] = |\Phi[x'_\perp(\sigma')]|^2
\ee
are then expressed as the square of Gaussian wavefunctionals~\cite{Brower:2006ea}, 
\be
\Phi[x_\perp(\sigma)] = \<x_\perp(\sigma)|0;0\> =  
\exp[ - \frac{1}{ 16 \pi^2 \alpha'} \oint d\sigma_1 \oint d\sigma_2 
\frac{x_\perp(\sigma_1) x_\perp(\sigma_2)}{\sin^2(\frac{\sigma_1 - \sigma_2}{2}) + \epsilon^2} ] \; ,
\ee
for the overlap of the string vacuum state, $|0;0\>$, and the string
bit distribution at the time of impact $x^+ = 0$.  Note that by the
state-operator correspondence~\cite{Polchinski:1998rq} the graviton
wave function also includes a factor $:\dd X^\mu(w) \bar \dd X^\nu(w )
\exp[i p X(w)]:$ at $|w| = 0$ (or $\tau = -\infty$) but this factor has
already been properly included in the spin-momentum factors for
each graviton external state in the planar amplitude (\ref{eq:planar}).

Thus we see that the geometrical extension of the transverse dimensions
that we saw above, where the KK radial mode $z$ allowed us to rewrite
a multi-channel problem in four dimensions using a transverse
$AdS_3$, has an analogue here.  For the string, the
exact flat space eikonal amplitude, a multi-channel problem involving
a tower of massive string states, is diagonalized using an infinite dimensional
space which is a product of
transverse impact-parameter spaces, one for each string bit.  During the
collision, each string bit interacts
instantaneously in light-cone time $X^+ = \tau$ undergoing {\it zero}
deflection.  The string bits are frozen.


\section{Unitarity, Confinement and Froissart Bounds}

In this section we address questions relating to unitarity,
confinement and the Froissart bound, in regimes where the eikonal
approximation in the bulk is believed to give the dominant
contribution to the field theory amplitude.  
If $b$ is sufficiently large compared to
$z,z'$, which indicate the sizes (at the moment of collision) of the scattering
objects, then the scattering interaction is weak and causes small
deflections.  In this limit the eikonal approximation is believed in some 
theories, including gravity, to be
a good estimate of the amplitude.

Our discussion below will be brief and we will not consider in detail
the effect of the hadron wave functions.  Instead we will just discuss
the bulk amplitudes at a given $z$ and $z'$.  This is a key input for
a computation of the full gauge theory amplitude. The wave functions
for hadron states peak near a value of $z$ corresponding to their
typical size.  The probability that a hadron fluctuates to a smaller
size ({\it i.e.}, is found at smaller $z$) is suppressed by a power
law. Meanwhile, the wave functions cut off very quickly at larger $z$.
Typically the bulk wave functions for hadrons have no support above
some maximal $z$; for example all hadrons in a confining theory are
cut off at some $z_{max}$, and a quarkonium state of mass $M$ has a
wave function with no support for $z>1/M$.  Thus, at large $b$ the
properties of the field theory amplitude are to a degree dominated by the
properties of the bulk eikonal amplitude at a particular $z$ and $z'$,
corresponding to the most likely sizes of the scattering hadrons.  However,
some of the physics can only be captured after integrating over
$z$ and $z'$.

At a given $z$ and $z'$, the cross-section for the partial wave
corresponding to $b$ approaches its unitarity bound when $|\chi|\sim
1$.  Since interactions become stronger at smaller $b$,
$\partial|\chi|/\partial b$ tends to be negative, so typically the
bound is reached for all $b$ less than some $b_{max}$, except possibly
for interference fringes.  If ${\rm Im}[\chi]>{\rm Re}[\chi]$, as is
the case for the weak-coupling Pomeron, the point $b_{max}$ is where
absorption becomes of order one, and one speaks of a black disk of
radius $b_{{\rm black}}$ where unitarity is saturated.  If the reverse
is true, as for the strong-coupling Pomeron, then outside the black
disk, whose radius is set by ${\rm Im}[\chi]\sim 1$, is a
``diffractive disk'', where one finds large average cross-sections
modulated by fringes.  The radius of this disk, $b_{{\rm diff}}$, is
set roughly by the condition ${\rm Re}[\chi]\sim 1$.  We emphasize
however that we are speaking of disks in the bulk, for fixed $z$, $z'$;
the corresponding disks in the gauge theory can be found only be 
integrating over $z$ and $z'$.

\subsection{Scattering in the Conformal Case}

Within a conformal theory on Minkowski space, there is no S-matrix for
the conformal modes themselves, but in the case of a large-$N$ gauge
theory we may add heavy quarks, build quarkonium states out of them,
and scatter these states off each other.  The technique of using
``onium-onium'' scattering to probe the (near-)conformal part of QCD
has a long history \cite{Mueller:1994gb}.  Quarkonium states have been
studied in AdS/CFT (see for example \cite{Karch:2002sh, myers}) and so
this study could be carried out in detail.  The calculation would
reduce to integrations over $z$ and $z'$ of the bulk eikonal
formula, weighted by the wave-functions of the onium states.

Preliminary to carrying out such a computation, we will focus some
attention on the properties of the bulk eikonal formula $-2i\widehat s
[e^{i\chi}-1]$ itself.  The parametric dependence of the various
physically interesting scales is quite intricate.  Their interplay,
and the physics for $\lambda$ closer to 1, deserves further
exploration than we will present here.

The kernel is obviously small
if $b$ is much larger than $z$ and $z'$, meaning that the two onium
states are far apart compared to their size.  Requiring the eikonal
phase be of order 1 tells us the radius $b_{{\rm diff}}$ where
diffraction sets in, for this value of $z,z'$.  Although the cut in
the $J$-plane dominates at any fixed $b,z,z'$ as $s$ becomes large,
the spin-2 exchange dominates at large $b$ for any fixed $s,z,z'$.  In
Eq.~(\ref{eq:onegraviton}), we saw
that the graviton exchange kernel is proportional to $G_3(2,v)\sim
1/b^6$ at very large $b$.  The condition ${\rm Re}[ \chi] \sim 1$
determines the radius of the diffractive disk, and if $\widehat s =
zz's \gg N^2$ and $b\gg z-z'$ this takes the form:
\be
b_{{\rm diff}}\sim \sqrt{zz'}\
 (zz's/N^2)^{1/6}  \ \ \ .
\ee
Since the graviton exchange is real, the disk has diffractive fringes and
is non-absorptive.  Note however that integrals over $z$ and $z'$ 
will wash out the fringes, giving full absorption.  This is interpretable
as due to the multi-channel
$2\to 2$ process discussed in Sec.~\ref{sec:multichannel}.

At a different radius, the effect of the higher-spin states becomes
important and the cut beginning at $j=j_0$ will dominate over the
spin-2 exchange.  Here we focus on the regime $\log s > \sqrt
\lambda/2$, which is where long-range effects from the diffusive
effect of the Pomeron can become important.  The transition between
the two regions occurs, from Eq. (\ref{eq:GPtransition}), at $v\sim
\widehat s^{2/\sqrt\lambda}$ (where $v$ is the chordal distance
defined in Eq.~(\ref{eq:vdefn}),) that is, for $z\approx z'$,
\be
b_{{\rm cross}} \sim  \sqrt{zz'}\ \sinh \log [(zz's)^{1/\sqrt\lambda}] \ .
\ee
At $b$ smaller than this, $\chi$ is determined by
Eq.~(\ref{eq:transvasympbeh}); its real part now grows slower than
$s^2$ and its imaginary part is nonzero due to $s$-channel production
of heavy hadrons.  We may now ask where ${\rm Im}[\chi]\sim
1$, using Eq.~(\ref{eq:transvasympbeh}).  For $\log s>\sqrt \lambda/2$,
this occurs at $\xi>1$ and thus $b/\sqrt{zz'}>1$.
For $z=z'$, the disk may become black at $b<b_{{\rm cross}}$, in which case
\be 
b_{{\rm black}} \sim 
 \sqrt{zz'}  \
\ \frac{(zz's)^{(j_0-1)/2}}{\lambda^{1/4}N}  \ \ ,
\ee
an expression which for self-consistency also requires
$s^{j_0-1}>\sqrt\lambda N^2$.  
But even though
the graviton exchange dominates ${\rm Re} [\chi]$ at 
$b>b_{{\rm cross}}$, the diffusive tail of the Pomeron can extend into
this region and dominate the imaginary part.  
It is in fact possible that $b_{{\rm black}}>b_{{\rm cross}}$,
in which case
\be 
b_{{\rm black}} \sim 
 \sqrt{zz'}  \
\left(\frac{ (zz' s)^{j_0-1}}{\lambda^{1/4} N}
\right)^{1/ \sqrt{2\sqrt\lambda}(j_0-1)}  \ ,
\ee
Again we emphasize that these formulas are at fixed $z$ and $z'$ with 
$z-z'\ll \sqrt{zz'} < b$.

Finally, let us recall that the Pomeron kernel is also small if $z$
and $z'$ are very different, even if $b=0$ (the case of a
color-transparent small onium passing through a large one).  Requiring
the eikonal phase for the Pomeron to be of order 1 gives a condition
which (for $z\ll z'$) is approximately
\be 
(z'/z) \sim (zz's/N^2)^{1/3} \ , 
\ee
where again it is the spin-2 exchange which dominates the result at
large $z'/z$.
Thus for any fixed
$z$ and $z'$, saturation takes over from color transparency as $s$
becomes large.  However, here one must treat the hadron wave functions
properly to obtain the full picture.

Of course, for these results to be correct, the eikonal approximation
must be valid.  The approximation breaks down if the scattering angle
is too large.  In the bulk coordinates this is the requirement that
$\partial \chi/\partial\sqrt{v} \ll \sqrt{\widehat s}$. 
For $v\gg
\widehat s^{2/\sqrt\lambda}\gg 1 $ this requires
\be
v > \frac{\widehat s^{1/7}}{N^{4/7}}
\ee
while for $\widehat s^{2/\sqrt\lambda}\gg v \gg 1$ it requires
\be
v > \frac{\widehat s^{\frac{2}{3}j_0-1}}{N^{4/3}} \ .
\ee
As written these equations are only valid if $v\sim e^\xi$, which
is only true if $v\gg 1$, and thus for self-consistency
they require (for $j_0\sim 2$) $s > N^4$.  For smaller $s$ the formulas
are modified.  Note that as $s$ becomes
very large compared to $N^4$, the right-hand side is larger than $b_{{\rm black}}$; there is no region, for $\lambda\gg 1$, where
the eikonal approximation is valid and the Pomeron cut is dominant.

We should note that these physical scales in position-space resemble
in many ways those found at $t=0$ in \cite{HIM}, where deep inelastic
scattering and saturation were analyzed.  This is because deep inelastic
scattering off an onium state, like an onium-onium scattering, probes 
the conformal regime.

\subsection{Confinement and the Froissart Bound}

Of course there is no Froissart bound in the conformal case 
because of long-range effects
in the conformal gauge theory.  If we want to see cross-sections
that grow like $(\log s)^2$ we need to turn to theories with confinement.

With confinement, we can discuss scattering of quarkonia or of
ordinary hadrons.  We will only touch on a few key points here,
leaving a more complete discussion for future work.  Also, we only
consider here the case where the beta function in the ultraviolet is
zero, and the $J$-plane has a cut at $j_0$; the physically relevant
case of a running coupling adds additional 
subtleties to an already complex subject.

Although we have spent most of this paper discussing the Pomeron in
strictly $AdS$ space, and our detailed formulas for the kernel (and in
particular, their simplicity) depend on conformal invariance of the
dual gauge theory, our methods generalize directly to
non-conformal cases.  It is completely straightforward to use the
$J$-plane, and to transform to transverse position space, in the case
of non-conformal gauge theories, including confining gauge theories.
The problems are purely technical.  Unfortunately, few non-conformal
cases are known that permit a largely analytic treatment even of the bulk
metric.  Only one seems approachable, the duality cascade \cite{KS}, and its
small {\it positive} beta function for the 't Hooft coupling
significantly changes the analytic structure in the $J$-plane.
Moreover, it appears likely that considerable model dependence
afflicts the small $t$ (and therefore large-$b$) behavior; also there
are some subtleties with identifying leading effects.  This
potentially means that the various analytically-tractable toy models
for confinement, including the hard-wall, D7-metric, and soft-wall
models \cite{Polchinski:2001tt,myers,Son:2003et,softwall}, are not
reliable here.  However, we will still be able to draw
some general conclusions.

\begin{figure}[h]
\begin{center}
\includegraphics[width=0.75\textwidth]{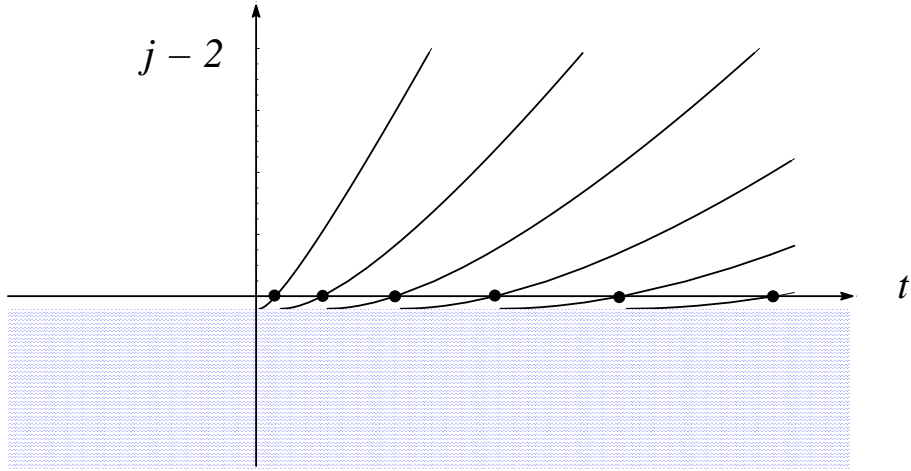}
\caption{The analytic behavior of  Regge trajectories in the hard-wall model,
  showing the location of the bound-state poles at $j = 2$ and the
  $t$-independent continuum cut (shaded) 
  at $j=j_0=2-2/\sqrt{\lambda}$ into which the Regge
  trajectories disappear.  The lowest Regge trajectory intersects the
  cut at a small positive value of $t$.  At sufficiently large $t$
  each trajectory attains a fixed slope, corresponding to the tension
  of the model's confining flux tubes.}
\label{fig:hardwall} 
\end{center}
\end{figure}

We begin with some general remarks.  In a confining theory,
the conformal kernel
${\cal K}(j,b^\perp,z,z')=(zz'/R^4)G_3(j,v)$ must be replaced with a more
complicated function, one which in a spectral representation should
exhibit a sum over discrete states and the presence of a mass gap in
the glueball spectrum.  In order to see how new scales can enter with
confinement, let us concentrate on the momentum-space Green's
function, ${\cal K}(j,t,z,z')$, which is the two-dimensional Fourier
transform of ${\cal K}(j,b^\perp,z,z')$, and can be obtained as the solution
to Eq.~(\ref{eq:ads5qDE1}), with $t=-q_\perp^2$. 

In the conformal limit, the $J$-plane consists simply of a single BFKL
cut at $j_0$.  We exhibited this using a spectral
representation in $J$, Eq.~(\ref{eq:jPomeron1}). Similarly, the lack
of a dimensionful scale leads to a continuous spectrum in $t$ beginning
at $t=0$.  Using a spectral representation in $t$, the kernel is
\be
{\cal K}(j,t,z,z') = \frac{(zz')^2}{2R^4}\int_0^\infty  dk^2 
\frac {J_{\widetilde \Delta(j)} (zk) J_{\widetilde \Delta(j)}(z'k)}{ k^2 -t  -i\epsilon } \label{eq:tspectral}
\ee
where  $\widetilde \Delta (j) \equiv \Delta_+(j)-2$, with $j>j_0$.
That is, ${\cal K}(j,t,z,z')$ has a branch cut along the positive
$t$-axis.

Confinement\footnote{For a confining theory,
the five-dimensional metric
$ds^2 = (R/z)^2 dz^2+ e^{-2A(z)} dx^2$ is
asymptotically $AdS_5$ ($e^{A(z)}\rightarrow z/R$ as $z\to 0$) and
is such that
for $z$ large,  $e^{-2A(z)}$ leads to an effective infrared
cutoff, $z<z_{max}$. Our main results for the eikonal representation,
Eqs.~(\ref{eq:adsiek1}-\ref{eq:ads3kernel1}), remain valid, after
appropriate kinematic modifications due to the confining
deformation. All explicit $z$ and $z'$ in various prefactors should be
replaced by $Re^{A(z)}$ and $Re^{A(z')}$ respectively, {\it e.g.}, $\widehat
s=zz' s$ becomes $\widehat s = R^2e^{A(z)+A(z')} s$. } 
leads to bound states in the
$t$-channel, with a discrete spectrum~~\cite{WittenAdS,Csaki:1998qr,deMelloKoch:1998qs,Aharony:1999ti,Constable:1999gb,Brower:1999nj,Brower:2000rp,PandoZayas:2003yb,Caceres:2005yx,Boschi-Filho:2005yh}.  The Green's function is now
given by a discrete sum of poles,
\be
{\cal K}(j,t,z,z') =e^{2A(z)} e^{2 A(z')} \sum_{n=0} \frac {\Phi_n(j,z) \Phi_n(j,z')} { t_n(j) -t  -i\epsilon } \;.
\ee
At $j=2$, these poles correspond to an infinite set of spin-two
glueballs.  We label these discrete modes sequentially,
$n=0,1,\cdots$, with the $t_0(j)$ pole interpolating the lightest
spin two glueball.\footnote{For each $n$, inverting $t_n(j)$ leads
to a Regge trajectory function, $\alpha_n(t)$. Due to a linear confining
potential, the trajectory functions are asymptotically
linear in $t$ at large $j$.}   In position space,
\be
{\cal K}(j,b,z,z') ={1\over 2\pi}e^{2A(z)} e^{2 A(z')} \sum_{n=0}  {\Phi_n(j,z) \Phi_n(j,z')} K_0(\sqrt{t_n(j)}b) \;.
\ee

Now we would like to understand the properties of ${\cal
K}(s,b,z,z')$, the inverse Mellin transform of the previous formula.
We focus on $z\sim z'\sim z_{max}$, which gives the largest
amplitude for fixed large $b$.
As in the conformal case, the nearby singularity at $j=2$ makes an
important contribution at very large $b$, for fixed $s$.
In this case the lightest spin-two glueball, with mass
$m_0 = \sqrt{t_0(2)}$ --- makes the most important contribution,
falling exponentially as $e^{-m_0b}/\sqrt{m_0 b}$.  At much smaller $b$,
again as in the conformal case,
one needs to account for various contributions to the discontinuity
across the cut in the $J$ plane starting at $j=j_0$.  This cut
reflects itself in the above formula in multiple ways.  First, the
wave functions $\Phi_n(j,z)$ are singular at $j=j_0$, with a square
root branch cut of order $\sqrt{ 2\sqrt{\lambda}(j-j_0)}$.  That this
is true is obvious from the fact that at small $b,z,z'$ our earlier
conformal result must be recovered.  Second, the trajectories $t_n(j)$
will in general have square root branch cuts, also of order $\sqrt{
2\sqrt{\lambda}(j-j_0)}$.

The hard-wall model, in which the metric is taken to be $AdS_5$ from
$z=0$ to $z=z_{max}$, and the space is cut off sharply at $z_{max}$,
illustrates these features.  This metric is not a solution to
the supergravity equations and has potentially problematic
non-analytic behavior at $z=z_{max}$, but it does realize confinement
and a mass gap, and it is analytically tractable.  The complete
$J$-plane structure of the hard-wall model, as well as the associated
kernel, was worked out in \cite{Brower:2006ea}, and is shown in
Fig.~\ref{fig:hardwall}; note in this model all the Regge trajectories
pass below the cut at positive $t$.  
The wave-functions at general $j$ are Bessel
functions, $\Phi_n(j,z) \sim J_{\widetilde\Delta(j)}(\sqrt{t_n(j)}z)$,
with the infinite set of discrete modes, $\{ t_n(j)\}$, determined by
boundary conditions at the infrared cutoff $z_{max}$.  These wave
functions pick up a square-root branch cut from the model-independent
cut in $\widetilde\Delta(j)=\Delta_+(j)-2$ at $j=j_0$; see
Eq.~(\ref{eq:Delta4}) for the definition of $\Delta_+(j)$.  Meanwhile,
the basic properties of the branch cuts of $t_n(j)$ in this model can
be inferred from Fig.~9 of \cite{Brower:2006ea}.

In general, to determine the full form of ${\cal K}(s,b,z,z')$
requires calculating the various contributions to the discontinuity
across the cut, and is not trivial.  But we also see that all of the
branch cuts are of order $\sqrt{ 2\sqrt{\lambda}(j-j_0)}$.  For $z,z'$ held fixed and of order $z_{max}$, the integration over $j-j_0$
will give a diffusion effect in $b$  of order 
$\exp[-c \sqrt \lambda b^2 m_0^2/\ln s]$, where $c$ is a constant of
order one that is model-dependent.
Thus the diffusive effect
in $b$ space extends only out to a distance proportional to
$(\lambda)^{-1/4} m_0^{-1}$.  The corresponding mass scale 
is associated with higher-spin hadrons that lie beyond the supergravity
regime.  Moreover, it appears that the leading trajectory typically
has the smallest discontinuity (this is certainly true in the hard
wall model) and thus gives the diffusive effect of largest range.

Now let us consider the effect of multiple scattering and unitarization. 
Since the effects of the Pomeron cut are short-range, the spin-2
poles dominate the physics at very large $b$ for fixed $s$ and
$z,z'\sim z_{max}$ (where the hadron wave functions are largest), with
the corrections from higher-spin states only becoming important at
shorter range.  Thus to understand the behavior of the cross-section,
we may focus on the spin-two glueball states.  Assuming only the
lightest glueball of mass $m_0$ is important, we find $|\chi| \sim 1$
inside a radius
\be\label{eq:bdiffX}
b_{{\rm diff}} \simeq  \frac{1}{m_0} \log (s/N^2\Lambda^2) + \dots
\ee
where $\Lambda \sim m_0$ is of order the light glueball masses.  
This approximation is self-consistent; the
contribution at this value of $b$ from the next-to-lightest glueball
state becomes relatively small as $s$ becomes large.

It is important to check whether the eikonal approximation is
self-consistent in the regimes we are discussing.  A weak but
necessary condition is that the scattering causes deflections at small
angle, which requires $b$ be larger than
\be\label{eq:bthetaX}
b_{\theta\ll 1} \sim  \frac{1}{2 m_0} \log (s/N^4\Lambda^2) + 
 \dots
\ee
The above formula is not quite right, as in this expression
we have assumed that only the lightest
glueball contributes, which is not true for moderately large $s$.  
But for our immediate purposes, it is enough that the above condition is valid
throughout the region where the lightest glueball dominates $\chi$,
and that the overlap of this region with the region $|\chi|>1$ has a
large area, proportional to $(\log [s/N^2\Lambda^2])^2$.  

In other words, the area in which the scattering amplitude is reaching its
unitarity bound, and in which the eikonal scattering is minimally
self-consistent, is of order $(\log s)^2$.  The coefficient of this
$(\log s)^2$ is bounded from above by the inverse mass-squared of the
lightest spin-2 glueball, and from below by an unknown (and
model-dependent) but nonzero coefficient.  This provides strong
evidence that the Froissart bound on the total cross-section
is not only satisfied, it is saturated.

Of course the eikonal approximation might break down at a radius
larger than that given by the above self-consistency condition.  But
unless this happens right at the edge of the diffractive disk, or our
formula for the eikonal phase quickly becomes a large overestimate,
the above argument that the Froissart bound is saturated remains
intact.  Moreover, on physical grounds, any changes to our formulas or
breakdown of the eikonal due to so-far unidentified effects are
unlikely to significantly weaken the scattering amplitude and bring
the amplitude below the unitarity bound in any of the region
$b<b_{{\rm diff}}$; in fact, the interactions being gravitational,
they are likely to make the scattering amplitude larger.  Thus our
conclusion appears robust.\footnote{We note that the proposal of
Giddings on the role of black holes and the Froissart bound
\cite{Giddings:2002cd,Kang:2004jd,Kang:2005bj} suggests but does not
strictly prove a lower bound.  Because of the difficulty of computing
the rate of black-hole production and the efficiency with which the
initial energy is converted to black hole mass, it is not clear to us
whether the lower-bound obtained from black hole production would be
larger or smaller than the one we are discussing here.  Note also that
because there might be other processes with larger cross-sections,
Giddings suggestion provides no upper bound.}

\section{Summary and Outlook}
\label{sec:disc}

In this paper, we have taken a step toward unitarization of high
energy scattering using string/gauge duality. The eikonal
approximation is a summation to all orders (in $1/N^2$, or $g_s$) of
multiple small-angle scatterings. Here we have computed scattering
amplitudes (or partial contributions to scattering amplitudes) in
large-$\lambda$ gauge theories by using the eikonal approximation for
multiple Pomeron exchange.  We have seen the required formalism is a
relatively straightforward generalization of our approach to multiple
graviton exchange in $AdS_5$ space. All we needed to do was convert
our earlier work on the Pomeron \cite{DIS, Brower:2006ea} from
momentum space to transverse position space, use a $J$-plane
representation of the amplitude, and combine it with the techniques of
\cite{Brower:2007qh}.

We carried this program out in its entirety in the case of a conformal
field theory, where the symmetries of the problem make it easy to
solve.  We showed that in transverse position space and the $J$-plane,
the Pomeron exchange amplitude is extremely simple: it is proportional
to a scalar $AdS_3$ propagator.  We examined the group-theoretic basis
of this result, comparing it to known results at weak coupling.  We
noted that the 
Pomeron cut dominates as $s$ goes to infinity for fixed $\lambda$, 
and recovered a graviton-exchange
kernel by holding $s$ fixed and letting $\lambda$ grow to infinity.
The eikonalization of this amplitude also had a number of interesting
features which we highlighted: a nontrivial phase compared to the
graviton, corresponding to production in the $s$-channel of excited
strings; a multi-channel interpretation; and a string-bit
interpretation.  These multiple viewpoints will be useful for the next
steps in the conformal case: corrections to the one-Pomeron exchange
approximation to the eikonal kernel from triple-Pomeron vertices, and
corrections beyond the eikonal approximation.  A further goal is a
complete Gribov-Regge effective theory in the large-$\lambda$ limit.

We finally turned to issues of unitarity in a bit more detail.  We
first considered how the Pomeron appears within the bulk amplitude
in the conformal case, noting where the graviton exchange
contribution takes over.  We also considered issues of color
transparency and the onset of saturation.  We then turned our
attention to confining theories.  Here we found unitarity saturated in
a disk with radius growing like $\log s$, given by multiple exchange
of light spin-two glueballs.  Within the eikonal approximation, it
appears that the Froissart bound is not only satisfied in the generic
large-$\lambda$ theory, it is also saturated.  To establish lower as
well as upper bounds on the cross-section in any given theory will
require more careful analysis.  

In future, it will be important to compute a variety of scattering
amplitudes and interpret the results; \cite{HIM} has recently begun
this program in the context of deep-inelastic scattering.  Eventually
one would hope to extract appropriate lessons for QCD, though this
will be a challenge, given the intricate dependence of the physics on
$s$, $b$, $\lambda$ and $N$.   In particular, the approach to the
region $\lambda\to 1$ holds some subtleties that are yet to be explored.

\noindent {\underline{Acknowledgments:}} 
We are pleased to acknowledge useful conversations
with D. Freedman, M. H. Fried, A. Kovner, J. Polchinski, and G. Veneziano.  The
work of R.C.B. was supported by the Department of Energy under
Contract.~No.~DE-FG02-91ER40676, that of M.J.S. by U.S. Department of Energy Contract.~No.~DE-FG02-96ER40956 and  that of C-I.T. was supported by the Department of Energy under Contract~No.~DE-FG02-91ER40688, Task-A.
C-I.T. and R.C.B. would like to thank the Aspen Center
for Physics for its hospitality during the writing of this paper.  We
are grateful to the Benasque Center for Science, where this work was
initiated.

\newpage
\bibliographystyle{utphys}
\bibliography{adspom}

\end{document}